\documentclass[journal]{IEEEtran}
\ifCLASSINFOpdf
\else
\fi

\usepackage[utf8]{inputenc}
\usepackage{amsmath,amssymb,amsfonts}
\usepackage{cite}

\usepackage{algorithmic}
\usepackage{stfloats}
\usepackage{graphicx}
\usepackage{textcomp}
\usepackage{xcolor}
\usepackage{color}
\usepackage{pifont}
\usepackage{amsthm}

\usepackage{hyperref}
\usepackage{mathrsfs}
\usepackage{booktabs}
\usepackage{hyperref}
\usepackage{cases}
\usepackage{comment}
\usepackage{empheq} 
\usepackage{caption}
\usepackage{subcaption}
\allowdisplaybreaks[4]

\usepackage{url}  
\usepackage{graphicx}  
\usepackage[ruled,vlined,linesnumbered]{algorithm2e}

\begin{document}
\captionsetup[figure]{labelfont={rm},labelformat={default},labelsep=period,name={Fig.}}
\let\sss= \scriptscriptstyle
\title{Latency Minimization for Movable Antennas-Enabled Relay-aided D2D Mobile Edge Computing Communication Systems}
%
%
%

\author{Yue Xiu,~Yang Zhao,~Ran Yang,~Huimin Tang,~Long Qu~\IEEEmembership{Senior~Member,~IEEE},~Maurice Khabbaz~\IEEEmembership{Senior~Member,~IEEE},
~Chadi Assi~\IEEEmembership{Fellow,~IEEE},
~Ning Wei~\IEEEmembership{Member,~IEEE}\\
\thanks{Yue Xiu, Ran Yang and Ning Wei are with 
National Key Laboratory of Science and Technology on Communications, University of Electronic Science and Technology of China, Chengdu 611731, China (E-mail:  
xiuyue12345678@163.com, yangran6710@outlook.com, wn@uestc.edu.cn).}
\thanks{Yang Zhao is with 
Nanyang Technological University, 50 Nanyang Ave, Singapore 639798, Singapore (email: zhao0466@e.ntu.edu.sg).}
\thanks{Huimin Tang is with the Cho Chun Shik Graduate School of Mobility, Korea Advanced Institute of Science and Technology, Daejeon, South Korea (email: tanghuimin1996@kaist.ac.kr).}
\thanks{L. Qu is with the Faculty of Electrical Engineering
 and Computer Science of Ningbo University, 315211 Ningbo, China. (email: qulongonline@gmail.com)}
\thanks{ 
 Maurice J. Khabbaz is with the CMPS Department, American University
 of Beirut, Lebanon (e-mail: mk321@aub.edu.lb).}
 \thanks{Chadi Assi is with Concordia University, Montreal, Quebec, H3G 1M8, Canada (E-mail:chadimassi@gmail.com).}
\thanks{The corresponding author is Ning Wei.}
}

\maketitle
\begin{abstract}
Device-to-device (D2D)-assisted mobile edge computing (MEC) is one of the critical technologies of future sixth generation (6G) networks. The core of D2D-assisted MEC is to reduce system latency for network edge UEs by supporting cloud computing services, thereby achieving high-speed transmission. Due to the sensitivity of communication signals to obstacles, relaying is adopted to enhance the D2D-assisted MEC system's performance and its coverage area. However, relay nodes and the base station (BS) are typically equipped with large-scale antenna arrays. This increases the cost of relay-assisted D2D MEC systems and limits their deployment. Movable antenna (MA) technology is used to work around this limitation without compromising performance. Specifically, the core of MA technology lies in optimizing the antenna positions to increase system capacity. Therefore, this paper proposes a novel resource allocation scheme for MA-enhanced relay-assisted D2D MEC systems. Specifically, the MA positions and beamforming of user equipments (UEs), relay, and BS as well as the allocation of resources and the computation task offloading rate at the MEC server, all are optimized herein with the objective of minimizing the maximum latency while satisfying computation and communication rate constraints. Since this is a multivariable non-convex problem, a parallel and distributed penalty dual decomposition (PDD) based algorithm is developed and combined with successive convex approximation (SCA) to solve this non-convex problem. The results of extensive numerical analyses show that the proposed algorithm significantly improves the performance of the MA-enhanced relay-assisted D2D communication system compared to a counterpart where relays and the BS are equiped with traditional fixed-position antenna (FPA).
\end{abstract}


%
\IEEEpeerreviewmaketitle

\section{Introduction}
\IEEEPARstart{I}{n} future sixth generation (6G) networks, developing internet of thing (IoT) devices such as drones and intelligent vehicles will significantly increase dependency on mobile data\cite{b1}. This surge in mobile data needs extensive storage and computation processing, which impacts system reliability and latency. Thus, future 6G networks must support ultra-reliable and low-latency communication (URLLC) to enable numerous applications\cite{b2}. The emergence of device-to-device (D2D)-assisted mobile edge computing (MEC) systems offers a technical scheme to address these challenges. The D2D-assisted MEC operates by introducing computation and data storage closer to where the network edge user equipments (UEs) need them\cite{b3}. Instead of transmitted data to a centralized cloud server, the D2D MEC processes data locally at edge nodes, such as base stations (BSs) or access points (APs). This approach reduces latency by improving bandwidth efficiency and offloading rate from the core network\cite{b4,b5}. However, due to the D2D offloading process, distance is also a crucial factor influencing energy consumption and latency. For intelligent devices, longer distances directly reduce the transmission rate and information quality, impacting energy consumption and latency during transmission. Relays effectively extend IoT coverage; hence, assisting UEs with offloading\cite{b6,b7}.

The literature comprises numerous work revolving around relay-assisted D2D MEC systems. \cite{b8} proposed a secure communication scheme for such systems involving flying eavesdroppers. \cite{b9} investigated a two-way relay network with multiple UE pairs with the objective of  maximizing the network-wide transmission rate without jeopardizing the quality of service (QoS). \cite{b10} explored servicing many UEs in the context of a multi-relay reflective intelligent surface (RIS) assisted communication environment. The authors aimed at minimizing the overall relay energy consumption and maintenance costs by jointly optimizing the RIS phase shifts. \cite{b11} investigated the problem of joint relay beamforming and resource allocation to minimize the latency of multi-relays-assisted computation offloading in an air-ground integrated millimeter-wave (mmWave) network. Therein, relays were equipped with functions allowing for ground D2Ds to access computationally empowered MEC servers. In this same context, a lower bound for the maximum achievable rate under low-to-medium signal-to-noise ratio (SNR) was derived in\cite{b12}, which also addressed the locally achieved edge-network latency. However, the surge in resource-intensive applications necessitates that the BS and relay employ multiple antennas to achieve a higher efficiency by using beamforming and fully exploiting the spatial degree of freedom (DoF). Traditional fixed-position antennas (FPAs) remain stationary; thus, fail to fully utilize the spatial DoF. Recently, movable antennas (MAs) were introduced to allow for the full DoF exploitation in continuous spatial regions\cite{b13}. Specifically, in MA-enhanced systems, each antenna is connected to the radio frequency (RF) chain via a flexible, length-adjustable cable. Each MA moves within the available spatial region\cite{b14}. Contrary to FPAs, MAs are flexibly/advantageously positioned to achieve better channel conditions and, thus, more efficient wireless communication. Generally, MA-enhanced systems utilize fewer antennas to leverage such spatial DoF, notably reducing the signal processing complexity and system cost compared to FPA-augmented systems \cite{b15}.

Many papers have already considered MA-enabled wireless communication systems For instance, the authors of \cite{b15} addressed traditional pre-coding in multiple MA-augmented UEs. They optimized antenna positions and the pre-coding matrix to minimize inter-user interference and transmission power. Similarly \cite{b16} optimized antenna positions to improve the transmission rate. In \cite{b17}, the authors explored MA-augmented cognitive radios to facilitate efficient spectrum sharing between primary and secondary users. The work of \cite{b18}, presented a general channel estimation framework for MA systems utilizing the multipath field response channel structure. Specifically, multiple channel measurements from Tx and Rx MA positions were collected and a compressed sensing method was employed to jointly estimate the angles of departure and arrival (AoD/AoA) as well as the complex multipath components' coefficients. \cite{b19} introduced an MA-aided multiple-input multiple-output (MIMO) system that optimizes the MA position to enhance capacity. Unlike traditional FPA MIMO systems, the proposed system flexibly adjusted the Tx/Rx MAs' positions, thereby reconfiguring the MIMO channels between them to achieve higher capacity.\\

\indent This paper aims at integrating MA in relay-assisted D2D MEC systems. Briefly, the potential of doing so is three-fold: \textit{a}) reducing the number of antennas in relay-assisted D2D MEC systems and, thus, lowering system costs, \textit{b}) introducing additional spatial DoFs; hence, enhancing resource allocation efficiency, and, \textit{c}) extending the coverage range of the BS and relay nodes. Yet, it is observed that the problem of jointly optimizing resource allocation, MA position and beamforming for the purpose of minimizing latency under resource scarcity and communication constraints, has been, if ever, very lightly addressed by existing publications (such as those surveyed above) in the context of real-world MA-enhanced relay assisted D2D MEC systems. Indeed, such a problem exhibits considerable challenges constituting a notable literature gap that this paper aims at closing. To this end, this paper's novel contributions and organization are summarized as follows:\\
%
%
\indent $1$) In Section \ref{II}, a novel MA-enabled relay-assisted D2D MEC system model is developed. Then, in Section \ref{III}, a joint resource allocation, MA position, and beamforming optimization problem is formulated to minimize the system-wide latency under resource scarcity and communication constraints.\\
\indent $2$) To work around the notable original problem's complexity (\textit{i.e.}, non-linearity and non-convexity), successive convex approximation (SCA) and penalty dual decomposition (PDD) are employed in Section \ref{IV} to guarantee convergence to a set of fixed solutions. Moreover, in this section, it is shown  that the SCA-and-PDD-based algorithms can be executed in parallel and their comprehensive complexity analysis is provided.\\
\indent $3$) In Section \ref{V}, extensive simulations are conducted to quantitatively gauge the merits of the MA-enhanced relay-assisted D2D MEC system as well as those of the newly proposed latency optimization framework. Reported results show that the proposed algorithm outperform existing counterparts by $\%14\sim\%35$. Finally, Section \ref{VI} concludes the paper.

\section{System Model and Problem Formulation}\label{II}

\begin{figure}
  \centering
  \includegraphics[scale=0.15]{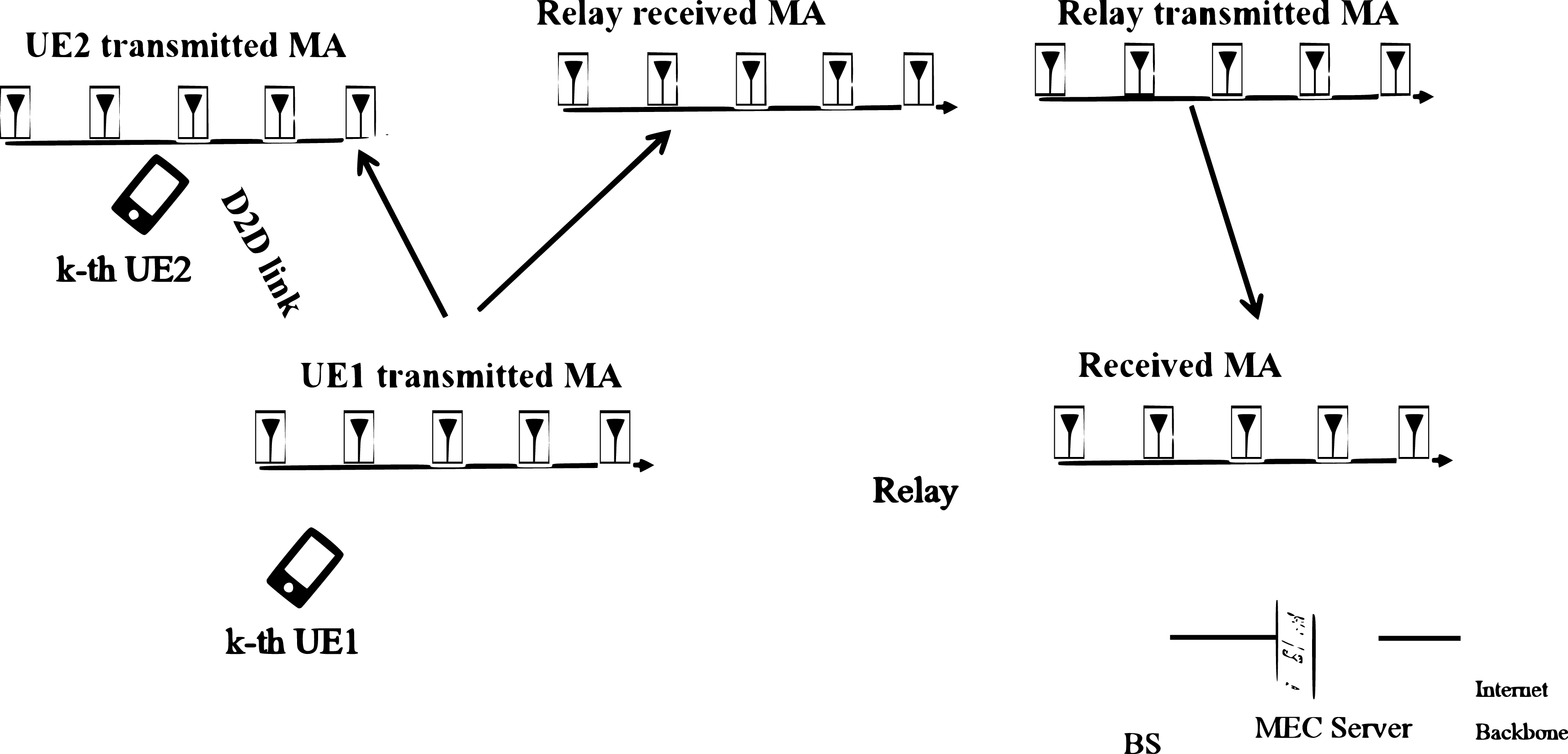}
  \captionsetup{justification=centering}
  \caption{Illustration of the relay-aided D2D MEC-based MAs.}
\label{FIGURE0}
\end{figure}
As shown in Fig.~\ref{FIGURE0}, the relay-assisted D2D MEC system supports MAs. This system consists of $K$ D2D pairs, each including two UEs, namely UE1 and UE2, each equipped with $N_{u}$ MAs, one relay with $N_{r}$ received MAs and $N_{t}$ transmitted MAs, a BS with $N_{b}$ MAs, and an MEC server. In this system, we assume the coordinate of the $k$-th UE1's $m$-th MA is represented as $\boldsymbol{t}_{k,m}=[x_{k,m},y_{k,m}]^{T}\in\mathcal{C}_{k, t}$, $m\in\{1,\ldots,N_{u}\}$, where $\mathcal{C}_{k, t}$ is the local movable region of the $k$-th UE1's MAs in the uplink transmission phase. In the D2D transmission phase, the coordinate of the $k$-th UE1's and UE2's $m$-th MA are represented as $\bar{\boldsymbol{t}}_{k,m}=[\bar{x}_{k,m},\bar{y}_{k,m}]^{T}\in\mathcal{C}_{k, t}$ and $\tilde{\boldsymbol{t}}_{k^{\prime},m}=[\tilde{x}_{k,m},\tilde{y}_{k,m}]^{T}\in\mathcal{C}_{k, t}$, respectively. The coordinate of the $n$-th received MA on the relay is denoted as $\boldsymbol{p}_{u,r,n}=[\boldsymbol{u}_{r,n}^{T}, h_{r}]^{T}\in\mathbb{R}^{3\times 1}$, where $h_{r}$ represents the relay height, and $\boldsymbol{u}_{r,n}=[x_{r,n},y_{r,n}]^{T}\in\mathcal{C}_{u,r}$, $n\in\{1,\ldots,N_{r}\}$ is the MA's azimuth coordinate. The coordinate of the $\tilde{n}$-th transmit MA on the relay is denoted as 
$\boldsymbol{p}_{u,t,\tilde{n}}=[\boldsymbol{u}_{t,\tilde{n}}^{T}, h_{t}]^{T}\in\mathbb{R}^{3\times 1}$, $\tilde{n}\in\{1,\ldots,N_{t}\}$, where $\boldsymbol{u}_{t,\tilde{n}}=[x_{t,\tilde{n}},y_{t,\tilde{n}}]^{T}\in\mathcal{C}_{u,t}$. The coordinates of the $j$-th antenna of the BS is denoted as $\boldsymbol{p}_{b,r,j}=[\boldsymbol{r}_{b,j}^{T}, h_{b}]^{T}\in\mathbb{R}^{3\times 1}$, where $h_{b}$ is the height of the BS. The areas $\boldsymbol{r}_{b,j}=[x_{b,j},y_{b,j}]^{T}\in\mathbb{R}^{2\times 1}\in\mathcal{C}_{b}$, $j\in\{1,\ldots,N_{b}\}$. $\mathcal{C}_{k, t}$,
$\mathcal{C}_{u,r}$, $\mathcal{C}_{u,t}$ and $\mathcal{C}_{b}$ represent the movement regions of the UE's MAs, the transmit and received MAs of the relay, and the BS's MAs, respectively.  Their expressions are in \cite{b15}.

\subsection{Channel model}
This paper assumes that the channels from the UE1 to the relay and from the relay to the BS are quasi-static block-fading channels. The channel can be updated for the uplink from the UE1 to the relay by adjusting the positions of the UE1's transmitted MAs and the relay's received MAs. The set of the coordinates for the transmitted MAs is denoted as $\boldsymbol{t}_{k}=[\boldsymbol{t}_{k,1},\boldsymbol{t}_{k,2},\ldots,\boldsymbol{t}_{k,N_{u}}]\in\mathbb{R}^{2\times N_{u}}$, and the set of coordinates for the received MAs is denoted as $\boldsymbol{u}_{r}=[\boldsymbol{u}_{r,1},\boldsymbol{u}_{r,2},\ldots,\boldsymbol{u}_{r,N_{r}}]\in\mathbb{R}^{2\times N_{r}}$. The channel from the $k$-th UE1 to the relay is represented as: 
\begin{align}
\boldsymbol{H}(\boldsymbol{t}_{k},\boldsymbol{u}_{r})=\boldsymbol{A}(\boldsymbol{u}_{r})^{H}\boldsymbol{\Sigma}_{k}\boldsymbol{A}(\boldsymbol{t}_{k})\in\mathbb{C}^{N_{r}\times N_{u}},\label{pro1}
\end{align}
where the field-response matrix (FRM) of the $k$-th UE is:
\begin{align}
\boldsymbol{A}(\boldsymbol{t}_{k})=\left[\boldsymbol{a}(\boldsymbol{t}_{k,m})\right]\in\mathbb{C}^{L\times N_{u}}, \forall~m\in\{1,\ldots,N_{u}\},\label{pro2}
\end{align}
in which $\boldsymbol{a}(\boldsymbol{t}_{k,m})=\left[e^{j\frac{2\pi}{\lambda}\kappa^{l}(\boldsymbol{t}_{k,m})}\right]^{T}\in\mathbb{C}^{L\times 1}$, $\forall~l\in\{1,\ldots,L\}$. $\lambda$ denotes the wavelength. Then, the difference in the signal propagation for the $l$-th transmit channel path between position $\boldsymbol{t}_{k,m}$ and the original coordinate is $\kappa^{l}(\boldsymbol{t}_{k,m})=(\boldsymbol{\pi}_{k}^{l})^{T}\boldsymbol{t}_{k,m}$, $\boldsymbol{\pi}_{k}^{l}=[\cos\varphi_{k}^{l}\cos\phi_{k}^{l},\cos\varphi_{k}^{l}\sin\phi_{k}^{l}]^{T}$, where $\phi_{k}^{l}$ is the azimuth AoD of the $l$-th transmit path is given by  $\phi_{k}^{l}\in[0,\pi]$, $\varphi_{k}^{l}\in[0,\pi]$. Similarly, $\boldsymbol{A}(\boldsymbol{u}_{r})$ is expressed as:
\begin{align}
\boldsymbol{A}(\boldsymbol{u}_{r})=[\boldsymbol{a}(\boldsymbol{u}_{r,n})]\in\mathbb{C}^{L\times N_{r}},\label{pro3}
\end{align}
where $\boldsymbol{a}(\boldsymbol{u}_{r,n})=\left[e^{j\frac{2\pi}{\lambda}\kappa^{l}(\boldsymbol{u}_{r,n})}\right]^{T}\mathbb{C}^{L\times 1}$, $\kappa^{l}(\boldsymbol{u}_{r,n})=(\boldsymbol{\pi}_{r}^{l})^{T}\boldsymbol{u}_{r,n}+h_{r}\sin\varphi_{r,n}^{l}$, $\boldsymbol{\pi}_{r}^{l}=[\cos\varphi_{r}^{l}\cos\phi_{r}^{l}, \cos\varphi_{r}^{l}\sin\phi_{r}^{l}]^{T}$, where $\phi_{r}^{l}$, $\varphi_{r}^{l}$ are the azimuth AoA of the $l$-th transmit path is given by $\phi_{r}^{l},\varphi_{r}^{l}\in[0,\pi]$. According to \cite{b13,b20}, since $\boldsymbol{H}(\boldsymbol{t}_{k},\boldsymbol{u}_{r})$ can be reconfigured and updated based on the MA positions, the channel is a function of both $\boldsymbol{t}_{k}$ and $\boldsymbol{u}_{r}$. Similarly, the channel from the relay to the BS, and the D2D channel can be represented as: 
\begin{align}
&\boldsymbol{G}(\boldsymbol{u}_{t},\boldsymbol{r}_{b})=\boldsymbol{A}(\boldsymbol{r}_{b})^{H}\tilde{\boldsymbol{\Sigma}}\boldsymbol{A}(\boldsymbol{u}_{t})\in\mathbb{C}^{N_{b}\times N_{t}},\nonumber\\
&\tilde{\boldsymbol{H}}(\bar{\boldsymbol{t}}_{k},\tilde{\boldsymbol{t}}_{k^{\prime}})=\boldsymbol{A}(\bar{\boldsymbol{t}}_{k})^{H}\tilde{\boldsymbol{\Sigma}}_{k,k^{\prime}}\boldsymbol{A}(\tilde{\boldsymbol{t}}_{k^{\prime}})\in\mathbb{C}^{N_{b}\times N_{t}},\label{pro4}
\end{align}
where $\boldsymbol{A}(\boldsymbol{u}_{t})$ and $\boldsymbol{A}(\boldsymbol{r}_{b})$ are denoted as:
\begin{align}
&\boldsymbol{A}(\bar{\boldsymbol{t}}_{k})=[\boldsymbol{a}(\bar{\boldsymbol{t}}_{k,m})]\in\mathbb{C}^{\bar{L}_{k,k^{\prime}}\times N_{u}},m\in\{1,\ldots,N_{u}\},\nonumber\\
&\boldsymbol{A}(\tilde{\boldsymbol{t}}_{k^{\prime}})=[\boldsymbol{a}(\tilde{\boldsymbol{t}}_{k^{\prime},m})]\in\mathbb{C}^{\bar{L}_{k,k^{\prime}}\times N_{u}},m\in\{1,\ldots,N_{u}\},\nonumber\\
&\boldsymbol{A}(\boldsymbol{u}_{t})=[\boldsymbol{a}(\boldsymbol{u}_{t,m})]\in\mathbb{C}^{\tilde{L}\times N_{t}},m\in\{1,\ldots,N_{t}\},\nonumber\\
&\boldsymbol{A}(\boldsymbol{r}_{b})=[\boldsymbol{a}(\boldsymbol{r}_{b,j})]\in\mathbb{C}^{\tilde{L}\times N_{b}},j\in\{1,\ldots,N_{b}\},\label{pro5}
\end{align}
where $\boldsymbol{a}(\tilde{\boldsymbol{t}}_{k,m})=\left[e^{j\frac{2\pi}{\lambda}\kappa_{t}^{\tilde{l}}(\tilde{\boldsymbol{t}}_{k,m})}\right]^{T}\in\mathbb{C}^{\bar{L}_{k,k^{\prime}}\times 1}$, $\boldsymbol{a}(\tilde{\boldsymbol{t}}_{k^{\prime},m})=$\\
$\left[e^{j\frac{2\pi}{\lambda}\kappa_{t}^{\tilde{l}}(\tilde{\boldsymbol{t}}_{k^{\prime},m})}\right]^{T}\in\mathbb{C}^{\bar{L}_{k,k^{\prime}}\times 1}$, $\boldsymbol{a}(\boldsymbol{u}_{t,m})=\left[e^{j\frac{2\pi}{\lambda}\kappa_{t}^{\tilde{l}}(\boldsymbol{u}_{t,m})}\right]^{T}\in$\\
$\mathbb{C}^{L\times 1}$ and $\boldsymbol{a}(\boldsymbol{r}_{b,j})=\left[e^{j\frac{2\pi}{\lambda}\kappa_{r}^{\tilde{l}}(\boldsymbol{r}_{b,j})}\right]^{T}\in\mathbb{C}^{L\times 1}$, $\tilde{l}\in\{1,\ldots,\tilde{L}\}$, $\bar{l}\in\{1,\ldots,\tilde{L}_{k,k^{\prime}}\}$, $\kappa^{\tilde{l}}(\boldsymbol{u}_{t,m})=(\boldsymbol{\pi}_{t}^{\tilde{l}})^{T}\boldsymbol{u}_{t,m}+h_{r}\sin\varphi_{t}^{\tilde{l}}$, $\boldsymbol{\pi}_{t}^{\tilde{l}}=[\cos\varphi_{t}^{\tilde{l}}\cos\phi_{t}^{\tilde{l}}, \cos\varphi_{t}^{\tilde{l}}\sin\phi_{t}^{\tilde{l}}]$, $\phi_{t}^{\tilde{l}},\varphi_{t,m}^{\tilde{l}}\in[0,\pi]$ and $\kappa_{r}^{\tilde{l}}(\boldsymbol{r}_{b,j})=(\boldsymbol{\pi}_{b}^{\tilde{l}})^{T}\boldsymbol{r}_{b,j}+h_{b}\sin\varphi_{b}^{\tilde{l}}$, $\boldsymbol{\pi}_{b}^{\tilde{l}}=[\cos\varphi_{b}^{\tilde{l}}\cos\phi_{b}^{\tilde{l}}, \cos\varphi_{b}^{\tilde{l}}\sin\phi_{b}^{\tilde{l}}]$, $\phi_{b}^{\tilde{l}},\varphi_{b}^{\tilde{l}}\in[0,\pi]$. $\kappa^{\tilde{l}}(\tilde{\boldsymbol{t}}_{k,m})=(\boldsymbol{\pi}_{k}^{\tilde{l}})^{T}\tilde{\boldsymbol{t}}_{k,m}+h_{r}\sin\vartheta_{k}^{\tilde{l}}$, $\boldsymbol{\pi}_{k}^{\tilde{l}}=[\cos\vartheta_{k}^{\tilde{l}}\cos\zeta_{k}^{\tilde{l}}, \cos\vartheta_{k}^{\tilde{l}}\sin\zeta_{k}^{\tilde{l}}]$, $\zeta_{k}^{\tilde{l}},\vartheta_{k,m}^{\tilde{l}}\in[0,\pi]$ and $\kappa_{r}^{\tilde{l}}(\tilde{\boldsymbol{k}}_{k^{\prime},m})=(\boldsymbol{\pi}_{k}^{\tilde{l}})^{T}\tilde{\boldsymbol{k}}_{k^{\prime},m}+h_{b}\sin\vartheta_{b}^{\tilde{l}}$, $\boldsymbol{\pi}_{k^{\prime}}^{\tilde{l}}=[\cos\vartheta_{k^{\prime}}^{\tilde{l}}\cos\zeta_{k}^{\tilde{l}}, \cos\vartheta_{k^{\prime}}^{\tilde{l}}\sin\zeta_{k^{\prime}}^{\tilde{l}}]$, $\zeta_{k^{\prime}}^{\tilde{l}},\vartheta_{k^{\prime}}^{\tilde{l}}\in[0,\pi]$.

\subsection{D2D Link Model}
According to the channel model in (\ref{pro4}), the received signal at the UE2 of the $k$-th D2D pair can be written as:
\begin{align}
&\boldsymbol{y}_{k}=\tilde{\boldsymbol{H}}(\bar{\boldsymbol{t}}_{k},\tilde{\boldsymbol{t}}_{k})\tilde{\boldsymbol{x}}_{k}+\sum\nolimits_{k^{\prime}=1,k^{\prime}\neq k}^{K}\tilde{\boldsymbol{H}}(\bar{\boldsymbol{t}}_{k},\tilde{\boldsymbol{t}}_{k^{\prime}})\tilde{\boldsymbol{x}}_{k^{\prime}}+\tilde{\boldsymbol{n}}_{r},\label{pro6}
\end{align}
where $\tilde{\boldsymbol{x}}_{k}=\tilde{\boldsymbol{w}}_{k}s_{k}\in\mathbb{C}^{N_{u}\times 1}$ is the transmitted signal from the UE2 in the $k$-th D2D link, and $\tilde{\boldsymbol{w}}_{k}\in\mathbb{C}^{N_{u}\times 1}$ represents the transmit beamforming for the D2D link. $\tilde{\boldsymbol{n}}_{r}\in\mathbb{C}^{N_{u}\times 1}\sim\mathcal{CN}(\boldsymbol{0},\tilde{\sigma}_{r}^{2}\boldsymbol{I})$ is the AWGN at the UE2. For the communication achievable rate at UE2 of the $k$-th D2D pair is expressed as:
\begin{align}
\tilde{C}_{k}=\log_{2}\left(1+\frac{\|\tilde{\boldsymbol{H}}(\bar{\boldsymbol{t}}_{k},\tilde{\boldsymbol{t}}_{k})\tilde{\boldsymbol{w}}_{k}\|^{2}}{\sum\nolimits_{k^{\prime}\neq k}\|\tilde{\boldsymbol{H}}(\bar{\boldsymbol{t}}_{k},\tilde{\boldsymbol{t}}_{k^{\prime}})\tilde{\boldsymbol{w}}_{k^{\prime}}\|^{2}+\tilde{\sigma}^{2}_{r}}\right).\label{pro7}
\end{align}

\subsection{Uplink Model}
According to the channel models in (\ref{pro1}) and (\ref{pro4}), the received signal processed by the relay can be written as
\begin{align}
\boldsymbol{y}_{r}=\sum\nolimits_{k=1}^{K}\boldsymbol{F}(\boldsymbol{H}(\boldsymbol{t}_{k},\boldsymbol{u}_{r})\boldsymbol{x}_{k}+\boldsymbol{n}_{r}),\label{pro8}
\end{align}
where $\boldsymbol{F}\in\mathbb{C}^{N_{t}\times N_{r}}$ is the relay beamforming matrix. $\boldsymbol{x}_{k}=\boldsymbol{w}_{k}s_{k}\in\mathbb{C}^{N_{u}\times 1}$ denotes the transmitted signal from the $k$-th UE1. $\boldsymbol{n}_{r}\in\mathbb{C}^{N_{r}\times 1}\sim\mathcal{CN}(\boldsymbol{0},\sigma_{r}^{2}\boldsymbol{I})$.  The received signal of the $k$-th UE1 after the processing at the relay is expressed as
\begin{align}
&\hat{s}_{k}=\boldsymbol{q}_{k}^{H}\boldsymbol{G}(\boldsymbol{u}_{t},\boldsymbol{r}_{b})\boldsymbol{y}_{r}=\boldsymbol{q}^{H}_{k}\boldsymbol{G}(\boldsymbol{u}_{t},\boldsymbol{r}_{b})\boldsymbol{F}\boldsymbol{H}(\boldsymbol{t}_{k},\boldsymbol{u}_{r})\boldsymbol{x}_{k}\nonumber+\boldsymbol{q}^{H}_{k}\\
&\sum\nolimits_{k^{\prime\prime}\neq k}^{K}\boldsymbol{G}(\boldsymbol{u}_{t},\boldsymbol{r}_{b})\boldsymbol{F}\boldsymbol{H}(\boldsymbol{t}_{k^{\prime\prime}},\boldsymbol{u}_{r})\boldsymbol{x}_{k^{\prime\prime}}+\boldsymbol{q}^{H}_{k}\boldsymbol{G}(\boldsymbol{u}_{t},\boldsymbol{r}_{b})\boldsymbol{F}\boldsymbol{n}_{r}\nonumber\\
&+\boldsymbol{q}^{H}_{k}\boldsymbol{n}_{b},\label{pro9}
\end{align}
where $\boldsymbol{Q}=[\boldsymbol{q}_{1},\ldots,\boldsymbol{q}_{k}]\in\mathbb{C}^{N_{b}\times K}$ denotes the receive beamforming matrix at the BS. $\boldsymbol{n}_{b}\in\mathbb{C}^{N_{b}\times 1}\sim\mathcal{CN}(\boldsymbol{0},\sigma_{b}^{2}\boldsymbol{I})$ is the AWGN at the BS. For the communication achievable rate at the $k$-th UE1 is expressed as formulation (\ref{pro10}).
\begin{figure*}
\begin{align}
C_{k}=\log_{2}\left(1+\frac{|\boldsymbol{q}_{k}^{H}\boldsymbol{G}(\boldsymbol{u}_{t},\boldsymbol{r}_{b})\boldsymbol{F}\boldsymbol{H}(\boldsymbol{t}_{k},\boldsymbol{u}_{r})\boldsymbol{w}_{k}|^{2}}{\sum\nolimits_{k^{\prime\prime}\neq k}|\boldsymbol{q}_{k}^{H}\boldsymbol{G}(\boldsymbol{u}_{t},\boldsymbol{r}_{b})\boldsymbol{F}\boldsymbol{H}(\boldsymbol{t}_{k^{\prime\prime}},\boldsymbol{u}_{r})\boldsymbol{w}_{k^{\prime\prime}}|^{2}+\|\boldsymbol{q}_{k}^{H}\boldsymbol{G}(\boldsymbol{u}_{t},\boldsymbol{r}_{b})\boldsymbol{F}\|^{2}\sigma^{2}_{r}+\|\boldsymbol{q}_{k}\|^{2}\sigma^{2}_{b}}\right).\label{pro10}
\end{align}
\hrulefill
\end{figure*}

\subsection{Computation Model}
\begin{figure}
  \centering
  \includegraphics[width=0.45\textwidth, height=0.19\textwidth]{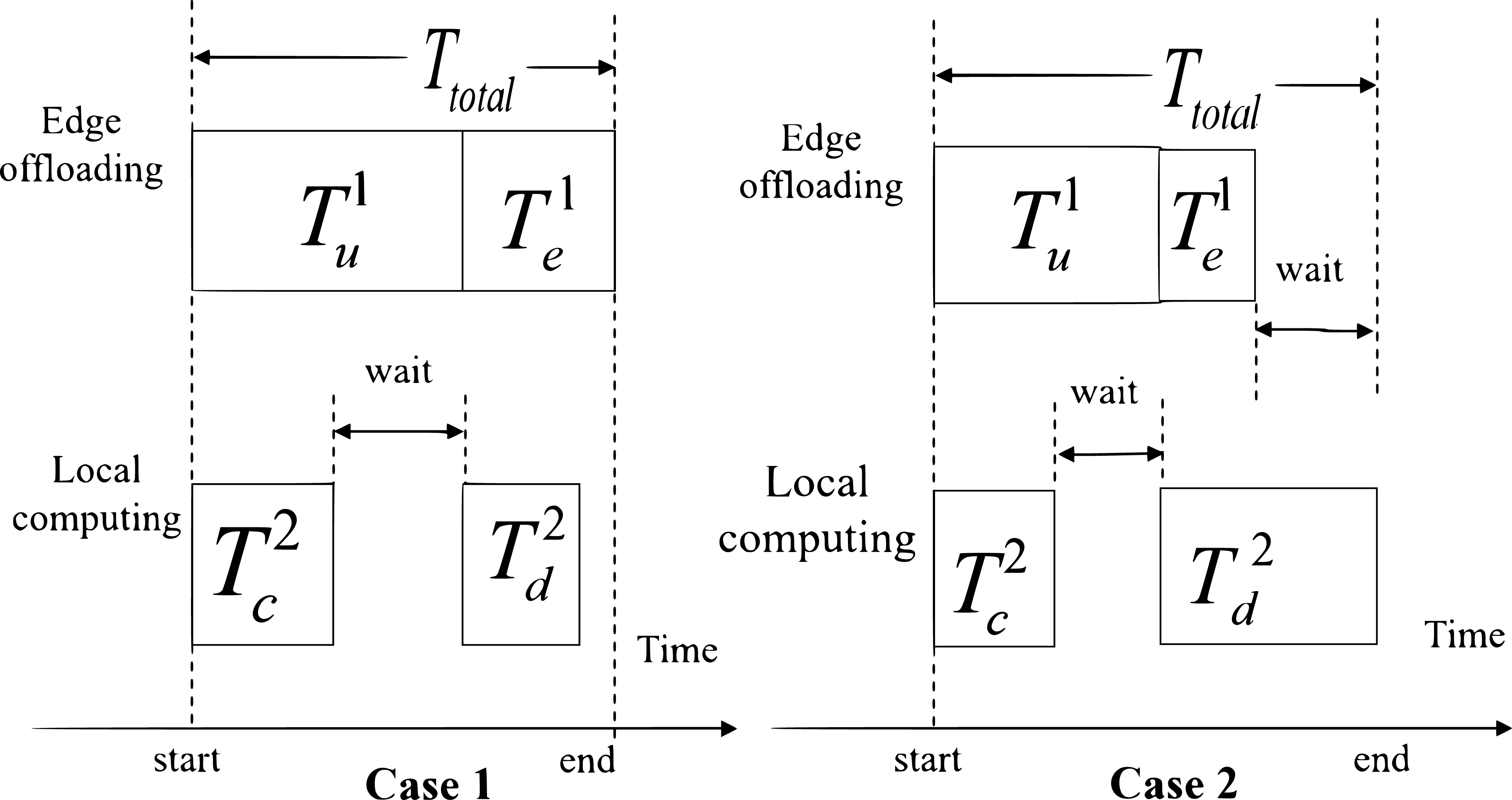}
  \captionsetup{justification=centering}
  \caption{Timeline of different offloading schemes.}
\label{FIGURE_0}
\end{figure}

In this MEC system model, UE1 has a total of $L_{a}$ computational tasks to be processed. These tasks are divided into two parts: one part of the $\rho L_{a}$ tasks is offloaded to the BS via a relay at an offloading rate of $\rho$ and processed by the MEC server, while the remaining $(1-\rho)L_{a}$ tasks are processed locally by UE1's CPU. The total offloading time includes the transmission time from UE1 to the relay then to the BS and the processing time at the MEC server. After completing the local computation, UE1 transmits the results to UE2 via the D2D link. Next, we adopt a general model\cite{b21}, where the computation results are represented as $\alpha L_{a}$ bits computational tasks, with $0\leq \alpha\leq 1$ representing the compression rate of the tasks. The value of $\alpha$ can be selected based on the task type and the algorithm used\cite{b21}. For convenience, we define $K_{L_{a}}=\frac{L_{a}}{F_{L_{a}}}$, $K_{E}=\frac{L_{a}}{F_{E}}$, $K_{1,k}=\frac{L_{a}}{C_{k}}$ and $K_{2,k}=\frac{\alpha L_{a}}{\tilde{C}}$, where $F_{L_{a}}$ and $F_{E}$ represent the local computation capability and edge computation capability (i.e., the MEC server's computation capability). The system latency can be divided into the following parts: local computation time: $T_{c}^{2} = (1-\rho) K_{L_{a}}$, MEC server computation time: $T_{e}^{1} = \rho K_{E}$, and offloading time from UE1 to the BS via the relay: $T_{u,k}^{1} = \rho K_{1,k}$, and the transmission latency from UE1 to UE2 for the computation results: $T_{d,k}^{2} = (1-\rho) K_{2,k}$. Let $T_{u}^{1}=\sum\nolimits_{k=1}^{K}T_{u,k}^{1}$ and $T_{d}^{2}=\sum\nolimits_{k=1}^{K}T_{d,k}^{2}$  denote the total offloading time and transmission latency. According to\cite{b22}, due to the introduction of the relay node to extend the communication system's coverage, the relay-assisted communication system has a longer transmission path, which increases the uplink transmission latency. Additionally, data must be transmitted to the relay node first and then from the relay to the BS, adding processing time compared to direct communication, further increasing the uplink transmission time. Therefore, in this paper, we assume $T_{u}^{1}\geq T_{c}^{2}$, and as shown in Fig.~\ref{FIGURE_0}, the specific process of computation and transmission is divided into two different scenarios.

When $T_{e}^{2}\geq T_{d}^{2}$ occurs, UE1's local computation is completed before task offloading. As a result, UE1 must wait for the offloading to be completed before transmitted the local results to UE2 via the D2D link. Furthermore, since the local results are transmitted before the edge computation, the BS can immediately send the edge computation results to UE2 without waiting for the D2D transmission. However, if $T_{e}^{2}<T_{d}^{2}$, UE1's local computation is still completed before task offloading, UE1 must again wait for the offloading to finish. In this case, however, the edge computation concludes before the local result transmission, requiring the BS to wait until the D2D link transmission is complete to avoid conflicts at UE2. Based on the two scenarios discussed above, the expression for the overall system latency is derived as follows
\begin{align}
T_{total}=T_{u}^{1}+\max\left\{T_{e}^{1},T_{d}^{2}\right\},\label{pro11}
\end{align}
Table.~\ref{TA1} summarizes the symbols used in this paper.

\begin{table}
\centering
\caption{Notations.}
\begin{tabular}{cp{3.8cm}}
\toprule[1pt]
\textbf{Symbol}&\textbf{Descriptions} \\
\midrule
$K$ & The number of D2D pairs. \\
$N_{u}/N_{r}/N_{t}/N_{b}$ &  The number of transmit/receive antennas of the UE, relay and BS. \\
$\boldsymbol{t}_{k,m}/\bar{\boldsymbol{t}}_{k,m}/\tilde{\boldsymbol{t}}_{k^{\prime},m}/\boldsymbol{u}_{r,n}$ & The coordinates of MAs. \\
$\boldsymbol{u}_{t,\tilde{n}}/\boldsymbol{r}_{b,j}$& The coordinates of MAs. \\
$\mathcal{C}_{k,t}/\mathcal{C}_{u,r}/\mathcal{C}_{u,t}/\mathcal{C}_{b}$ & MA mobile region. \\
$\tilde{\boldsymbol{H}}(\bar{\boldsymbol{t}}_{k},\tilde{\boldsymbol{t}}_{k^{\prime}})/\boldsymbol{H}(\boldsymbol{t}_{k},\boldsymbol{u}_{r})/\boldsymbol{G}(\boldsymbol{u}_{t},\boldsymbol{r}_{b})$ & The Channel between UE and relay/between relay and BS/between D2D pair. \\
$L_{a}$ & Number of computational tasks. \\
$L/\tilde{L}/\bar{L}_{k,k^{\prime}}$ & The number of channel paths. \\
$\boldsymbol{A}(\boldsymbol{t}_{k})/\boldsymbol{A}(\boldsymbol{u}_{t})/\boldsymbol{A}(\bar{\boldsymbol{t}}_{k})$ & Tx MA field response matrix. \\
$\boldsymbol{A}(\boldsymbol{u}_{r})/\boldsymbol{A}(\boldsymbol{r}_{b})/\boldsymbol{A}(\tilde{\boldsymbol{t}}_{k^{\prime}})$ & Rx MA field response matrix. \\
$\varphi_{k}^{l}/\varphi_{r}^{l}/\varphi_{t}^{\tilde{l}}/\varphi_{b}^{\tilde{l}}/\zeta_{k}^{\tilde{l}}$ & The AoAs. \\
$\phi_{k}^{l}/\phi_{r}^{l}/\phi_{t}^{\tilde{l}}/\phi_{b}^{\tilde{l}}/\vartheta_{k}^{\tilde{l}}$ & The AoDs. \\
$\boldsymbol{F}/\boldsymbol{Q}/\boldsymbol{w}_{k}/\tilde{\boldsymbol{w}}_{k}$ & The Tx/Rx beamforming. \\
$\boldsymbol{n}_{r}/\tilde{\boldsymbol{n}}_{r}/\boldsymbol{n}_{b}$ & The Noise signal. \\
$T_{k}$ & Number of computation task bits. \\
$\alpha$ & Task compression ratio. \\
$\rho$ & The offloading ratio. \\
$T_{u}^{1}/T_{e}^{1}/T_{c}^{2}/T_{d}^{2}$ & Offloading time from UE1 to the BS/ computing time at the MEC server/ local computing time/latency for transmit the computational result from
UE1 to UE2. \\
$P_{r}/P_{k}$ & The relay/UE1 transmit power. \\
$K_{L_{a}}/K_{E}/K_{1,k}/K_{2,k}$ &  The computing time/computing time at the MEC server/time from UE1 to the BS/latency for transmit from
the BS to UE2.\\
$F_{L_{a}}/F_{E}$ & The local computational capability/edge computational capability. \\
$C_{k}/\tilde{C}_{k}$ & The system capacity. \\
$T_{total}$ & The overall system latency. \\
\bottomrule[1pt]
\end{tabular}\label{TA1}
\end{table}

\subsection{Problem Formulation}
In the proposed MA-enable relay-aided D2D MEC system, we focus on the joint design of several key parameters to minimize the maximum total system latency $T_{total}$. Specifically, we aim to minimize the system latency by optimizing the beamforming vectors, offloading rates, and the position of the MA. Therefore, the system latency minimization problem can be mathematically formulated as follows:
\begin{subequations}
\begin{align}
\min_{\mathcal{U}}&~T_{total},\label{pro12a}\\
\mbox{s.t.}~
&T_{u}^{1}\geq T_{c}^{2},&\label{pro12b}\\
&\|\boldsymbol{w}_{k}\|^{2}\leq P_{k}, \forall~k=1,\ldots,K,&\label{pro12c}\\
&\|\tilde{\boldsymbol{w}}_{k}\|^{2}\leq P_{k}, \forall~k=1,\ldots,K,&\label{pro12d}\\
&\|\boldsymbol{F}\|^{2}\leq P_{r},&\label{pro12e}\\
&0\leq \rho\leq 1, &\label{pro12f}\\
&\boldsymbol{v}_{\ell}\in\mathcal{C}_{\ell, v},&\label{pro12g}\\
&\|\boldsymbol{v}_{\ell,o_{1}}-\boldsymbol{v}_{\ell,o_{2}}\|_{2}\geq D,  o_{1}\neq o_{2},&\label{pro12h}
\end{align}\label{pro12}%
\end{subequations}
in which $\boldsymbol{v}_{\ell}\in\{\boldsymbol{t}_{k}$,$\tilde{\boldsymbol{t}}_{k^{\prime}}$,$\bar{\boldsymbol{t}}_{k}$,$\boldsymbol{u}_{r}$,$\boldsymbol{u}_{t}$,$\boldsymbol{r}_{b}\}$, $\ell\in\{k$,$k^{\prime}$,$r$,$t$,$b\}$, $\mathcal{C}_{\ell, v}\in\{\mathcal{C}_{k, t}$,$\mathcal{C}_{k,\bar{t}}$,$\mathcal{C}_{u,r}$,$\mathcal{C}_{u,t}$,$\mathcal{C}_{b}\}$, $o_{1}\in\{m_{1}$,$n_{1}$,$\tilde{n}_{1}$,$j_{1}\}$, $o_{2}\in\{m_{2},n_{2}$,$\tilde{n}_{2},j_{2}\}$. $\mathcal{U}=\{\boldsymbol{F}$,$\boldsymbol{v}_{\ell}$,$\rho$,$\boldsymbol{Q}$,$\boldsymbol{w}_{k}$, $\tilde{\boldsymbol{w}}_{k}\}$. The constraints (\ref{pro12b}) and (\ref{pro12c}) denote the transmit power constraints of D2D UEs. (\ref{pro12e}) is the transmit power constraint of the relay.  (\ref{pro12g}) is the  MA moving region of the UEs, relay, and BS.  (\ref{pro12h}) is to avoid the coupling effect between the antennas in the transmit/receive region, and $D$ is the minimum distance between each pair of antennas. Problem (\ref{pro12}) is challenging to solve due to the non-convexity of the channel vectors and achievable rates with respect to the positions of MAs. Additionally, the coupling of these high-dimensional matrix and vector variables further complicates the problem. Existing optimization tools cannot provide an optimal solution for (\ref{pro12}) within polynomial time. Therefore, we develop suboptimal solutions for (\ref{pro12}) in the following section.

\section{Proposed Joint Resource Allocation Algorithm}\label{III}
In this section, we reformulate problem (\ref{pro12}) into a more tractable equivalent form. We apply the PDD method to address the highly coupled terms, introducing auxiliary variables and equality constraints to simplify the problem into one with multiple separable equality constraints. These constraints are then penalized and incorporated into the objective function, resulting in an augmented lagrangian (AL) problem \cite{b23,b24}. The proposed PDD method involves two iterative loops: in the inner loop, we solve the AL problem using an efficient algorithm based on CCCP\cite{b25,b26,b27} within a BCD framework; in the outer loop, we adjust the dual variables or penalty parameters according to the level of constraint violation. Finally, we summarize the proposed algorithm and assess its computational complexity.

To address problem (\ref{pro12}), we first introduce the following auxiliary variables: $t_{u}$, $t_{u,k}$,$t_{d}$, $t_{d,k}$, $t_{c}$, and $t_{c,k}$, $\forall~k\in\mathcal{K}$. We define $t_{d}=\max\{T_{e}^{1}, \sum_{k=1}^{K}T_{d,k}^{2}\}$, which implies that $t_{d}\leq T_{e}^{1}$ and $t_{d}\leq \sum_{k=1}^{K}T_{d,k}^{2}$. Subsequently, we set 
$t_{d,k}\leq T_{d,k}^{2}$ and $t_{d}=\sum_{k=1}^{K}t_{d,k}$, allowing $t_{d}\leq \sum_{k=1}^{K}T_{d,k}^{2}$ to be equivalently transformed into $t_{d,k}\leq T_{d,k}^{2}$ and $t_{d}=\sum_{k=1}^{K}t_{d,k}$. Similarly, we set $T_{u,k}^{1}\leq t_{u,k}$ and $t_{u}=\sum_{k=1}^{K}t_{u,k}$. Finally, based on the relationship $T_{u}^{1}\geq T_{c}^{2}$, we introduce an auxiliary variable $t_{c}$and enforce $T_{u}^{1}\geq t_{c}\geq T_{c}^{2}$. Since 
$T_{u}^{1}=\sum_{k=1}^{K}t_{u,k}\geq t_{c}$ remains challenging to handle directly. We further introduce the auxiliary variables $t_{c,k}$ and set $t_{c}=\sum_{k=1}^{K}t_{c,k}$ and $t_{u,k}\geq t_{c,k}$. Based on the above, the problem is reformulated as:
\begin{subequations}
\begin{align}
\min_{\hat{\mathcal{U}}}&~t_{u}+t_{d},\label{pro13a}\\
\mbox{s.t.}~
&T_{u,k}^{1}\leq t_{u,k}, T_{e}^{1}\leq t_{d},T_{d,k}^{2}\leq t_{d,k},T_{u,k}^{1}\geq t_{c,k},&\label{pro13b}\\
&t_{d}=\sum_{k=1}^{K}t_{d,k},t_{u}=\sum_{k=1}^{K}t_{u,k},t_{c}=\sum_{k=1}^{K}t_{c,k},&\label{pro13c}\\
&(\ref{pro12c})-(\ref{pro12h}),&\label{pro13d}
\end{align}\label{pro13}%
\end{subequations}
where $\hat{\mathcal{U}}=\{$$t_{u}$, $t_{d}$, $t_{c}$, $t_{u,k}$, $t_{d,k}$, $t_{c,k}$, $\boldsymbol{F}$, $\boldsymbol{v}_{\ell}$, $\rho$, $\boldsymbol{Q}$, $\boldsymbol{w}_{k}$, $\tilde{\boldsymbol{w}}_{k}\}$. Problem (\ref{pro13}) is rewritten as:
\begin{subequations}
\begin{align}
\min_{\hat{\mathcal{U}}}&~t_{u}+t_{d},\label{pro14a}\\
\mbox{s.t.}~
&\rho L_{a}/C_{k}\leq t_{u,k}, &\label{pro14b}\\
&\rho L_{a}/F_{E}\leq t_{d},&\label{pro14c}\\
&\alpha(1-\rho)L_{a}/\tilde{C}_{k}\leq t_{d,k},&\label{pro14d}\\
&\rho L_{a}/C_{k}\geq t_{c,k},&\label{pro14e}\\
&(\ref{pro12c})-(\ref{pro12h}),(\ref{pro13b}),(\ref{pro13c})&\label{pro14f}
\end{align}\label{pro14}%
\end{subequations}
Then, in the proposed system, the SINR expressions (\ref{pro7}) and (\ref{pro10}) are complex and take a fractional form, making it highly challenging to solve the associated constraint directly. Specifically, the signal, interference, and noise power in the SINR expression are determined by multiple coupled variables, further complicating the SINR expression's handling. To address this problem, we introduce auxiliary variables $\eta_{k}$ and $\tilde{\eta}_{k}$. These auxiliary variables facilitate the simplification of the original SINR expression. Specifically, the introduction of auxiliary variables $\eta_{k}$ and $\tilde{\eta}_{k}$ allows the complex fractional form to be re-expressed in a more manageable format, and $C_{k}$ and $\tilde{C}_{k}$ are re-expressed as:
\begin{align}
&C_{k}\leq\log_{2}(1+\eta_{k}),\tilde{C}_{k}\leq\log_{2}(1+\tilde{\eta}_{k}),\label{pro15}
\end{align}
in which $\eta_{k}$ and $\tilde{\eta}_{k}$ are given 
in (\ref{pro16}) at the top of next page and (\ref{pro17}).
\begin{figure*}
\begin{align}
\eta_{k}\leq \frac{|\boldsymbol{q}_{k}^{H}\boldsymbol{G}(\boldsymbol{u}_{t},\boldsymbol{r}_{b})\boldsymbol{F}\boldsymbol{H}(\boldsymbol{t}_{k},\boldsymbol{u}_{r})\boldsymbol{w}_{k}|^{2}}{\sum\nolimits_{k^{\prime\prime}\neq k}|\boldsymbol{q}_{k}^{H}\boldsymbol{G}(\boldsymbol{u}_{t},\boldsymbol{r}_{b})\boldsymbol{F}\boldsymbol{H}(\boldsymbol{t}_{k^{\prime\prime}},\boldsymbol{u}_{r})\boldsymbol{w}_{k^{\prime\prime}}|^{2}+\|\boldsymbol{q}_{k}^{H}\boldsymbol{G}(\boldsymbol{u}_{t},\boldsymbol{r}_{b})\boldsymbol{F}\|^{2}\sigma^{2}_{u}+\|\boldsymbol{q}_{k}\|^{2}\sigma^{2}_{b}}.\label{pro16}
\end{align} 
\hrulefill
\end{figure*}
\begin{align}
\tilde{\eta}_{k}\leq \frac{\|\tilde{\boldsymbol{H}}(\bar{\boldsymbol{t}}_{k},\tilde{\boldsymbol{t}}_{k})\tilde{\boldsymbol{w}}_{k}\|^{2}}{\sum\nolimits_{k^{\prime}\neq k}\|\tilde{\boldsymbol{H}}(\bar{\boldsymbol{t}}_{k},\tilde{\boldsymbol{t}}_{k^{\prime}})\tilde{\boldsymbol{w}}_{k^{\prime}}\|^{2}+\sigma^{2}_{d}}.\label{pro17}
\end{align}
To cope with the coupling constraints shown in (\ref{pro14b})-(\ref{pro14e}),  (\ref{pro16}) and (\ref{pro17}), we
further introduce a set of auxiliary variables and they are given in the follows table
\begin{table}
\centering
\caption{Auxiliary variables.}
\begin{tabular}{cp{4.2cm}}
\toprule[1pt]
$C_{k}=\hat{C}_{k}=\bar{C}_{k}$, $\tilde{C}_{k}=\check{C}_{k}$, $\rho=\hat{\rho}=\tilde{\rho}=\bar{\rho}=\check{\rho}$, $t_{d}=\hat{t}_{d}=\tilde{t}_{d}$.\\
$t_{u}=\tilde{t}_{u}$, $t_{d,k}=\tilde{t}_{d,k}$, $t_{c,k}=\tilde{t}_{c,k}$, $t_{u,k}=\tilde{t}_{u,k}$, $\eta_{k}=\tilde{\eta}_{k}$, $\hat{\eta}_{k}=\bar{\eta}_{k}$.\\
$\boldsymbol{F}=\tilde{\boldsymbol{F}}$, $\hat{\boldsymbol{w}}_{k}=\boldsymbol{w}_{k}$, $\bar{\boldsymbol{w}}_{k}=\tilde{\boldsymbol{w}}_{k}$,
$\tilde{u}_{k,k^{\prime\prime}}=\tilde{\boldsymbol{u}}_{k}\tilde{\boldsymbol{\mu}}_{k^{\prime\prime}}$.\\ $\tilde{\boldsymbol{u}}_{k}=\tilde{\boldsymbol{v}}_{b,k}\boldsymbol{F}$, $\tilde{\boldsymbol{v}}_{b,k}=\boldsymbol{v}_{b,k}^{H}\tilde{\boldsymbol{\Sigma}}\boldsymbol{B}_{t}$, $\tilde{\boldsymbol{\mu}}_{k^{\prime\prime}}=\boldsymbol{B}_{r}^{H}\boldsymbol{\Sigma}_{k}\boldsymbol{\mu}_{k^{\prime\prime}}$, $\tilde{\boldsymbol{\mu}}_{k^{\prime\prime}}=\boldsymbol{B}_{r}^{H}\boldsymbol{\Sigma}_{k}$\\
$\times\boldsymbol{\mu}_{k^{\prime\prime}}$. $\tilde{\boldsymbol{\omega}}_{k,k^{\prime}}=\boldsymbol{B}_{\bar{t},k}\tilde{\boldsymbol{\Sigma}}_{k,k^{\prime}}\boldsymbol{\omega}_{k^{\prime}}$, $\boldsymbol{\omega}_{k^{\prime}}=\boldsymbol{B}_{\tilde{t},k^{\prime}}\tilde{\boldsymbol{
w}}_{k^{\prime}}$, $\boldsymbol{\mu}_{k^{\prime\prime}}=\boldsymbol{B}_{t,k^{\prime\prime}}\tilde{\boldsymbol{w}}_{k^{\prime\prime}}$.\\
$\boldsymbol{B}_{t,k}=\boldsymbol{A}(\boldsymbol{t}_{k})$, $\boldsymbol{B}_{t}=\boldsymbol{A}(\boldsymbol{u}_{t})$, $\boldsymbol{B}_{r}=\boldsymbol{A}(\boldsymbol{u}_{r})$, $\boldsymbol{B}_{b}=\boldsymbol{A}(\boldsymbol{r}_{b})$.\\ $\boldsymbol{B}_{\bar{t},k}=\boldsymbol{A}(\bar{\boldsymbol{t}}_{k})$, $\boldsymbol{B}_{\tilde{t},k^{\prime}}=\boldsymbol{A}(\tilde{\boldsymbol{t}}_{k^{\prime}})$. \\
$|\boldsymbol{B}_{t,k}(l,m_{1})|=1$, $|\boldsymbol{B}_{r}(l,n_{1})|=1$, $|\boldsymbol{B}_{t}(\tilde{l},\tilde{n}_{1})|=1$, $|\boldsymbol{B}_{b}(\tilde{l},j_{1})|=1$.\\
$|\boldsymbol{B}_{\tilde{t},k^{\prime}}(l_{k,k^{\prime}},m_{1})|=1$,$|\boldsymbol{B}_{\tilde{t},k}(l_{k,k^{\prime}},m_{1})|=1$. \\
\bottomrule[1pt]
\end{tabular}\label{TA2}
\end{table} 

The movement of antennas can cause the relationship between antenna positions and the array response matrix to become nonlinear. To decouple this complex nonlinear relationship, we introduce an auxiliary variable 
$\boldsymbol{B}_{t,k},\tilde{\boldsymbol{B}}_{t}, \boldsymbol{B}_{r}, \boldsymbol{B}_{b}, \boldsymbol{B}_{\bar{t},k}, \boldsymbol{B}_{\tilde{t},k^{\prime}}$. Given that all elements in the array response matrix must satisfy a constant modulus one constraint, we consider transforming the auxiliary variables $\boldsymbol{B}_{t,k},\tilde{\boldsymbol{B}}_{t}, \boldsymbol{B}_{r}, \boldsymbol{B}_{b}, \boldsymbol{B}_{\bar{t},k}, \boldsymbol{B}_{\tilde{t},k^{\prime}}$ into a constant modulus-constrained auxiliary variable. 
Therefore:
\begin{subequations}
\begin{align}
\min_{\tilde{\mathcal{U}}}&~t_{u}+t_{d},\label{pro18a}\\
\mbox{s.t.}~
&\bar{\rho} L_{a}/\bar{C}_{k}\leq\tilde{t}_{u,k},\hat{\rho} L_{a}/F_{E}\leq\tilde{t}_{d},&\label{pro18b}\\
&\alpha(1-\tilde{\rho}) L_{a}/\check{C}_{k}\leq\tilde{t}_{d,k},\check{\rho} L_{a}/\hat{C}_{k}\leq\tilde{t}_{c,k},&\label{pro18c}\\
&C_{k}\leq\log_{2}(1+\eta_{k}),\tilde{C}_{k}\leq\log_{2}(1+\hat{\eta}_{k}),&\label{pro18d}\\
&\hat{t}_{d}=\sum\nolimits_{k=1}^{K}t_{d,k},\tilde{t}_{u}=\sum\nolimits_{k=1}^{K}t_{u,k}, t_{c}=\sum\nolimits_{k=1}^{K}t_{c,k},&\label{pro18e}\\
&0\leq\rho\leq 1,&\label{pro18f}\\
&\sum_{k^{\prime\prime}\neq k}|\tilde{u}_{k,k^{\prime\prime}}|^{2}+\|\tilde{\boldsymbol{u}}_{k}\|^{2}\sigma_{r}^{2}+\|\boldsymbol{q}_{k}\|^{2}\sigma_{b}^{2}-\frac{|\tilde{u}_{k,k}|^{2}}{\tilde{\eta}_{k}}\leq 0,&\label{pro18g}\\
&\sum_{k^{\prime}\neq k}\|\tilde{\boldsymbol{\omega}}_{k,k^{\prime}}\|^{2}+\sigma_{b}^{2}-\frac{\|\tilde{\boldsymbol{\omega}}_{k,k}\|^{2}}{\bar{\eta}_{k}}\leq 0,&\label{pro18h}\\
&\|\tilde{\boldsymbol{F}}\|^{2}\leq P_{r}, \|\boldsymbol{w}_{k}\|^{2}\leq P_{k}, \|\tilde{\boldsymbol{w}}_{k}\|^{2}\leq P_{k},&\label{pro18i}\\
&\boldsymbol{v}_{\ell}\in\mathcal{C}_{\ell, v},&\label{pro18j}\\
&\tilde{\boldsymbol{v}}_{\ell,o_{1}}=\boldsymbol{v}_{\ell,o_{1}}-\boldsymbol{v}_{\ell,o_{2}}, \|\tilde{\boldsymbol{v}}_{\ell,o_{1}}\|_{2}\geq D,  o_{1}\neq o_{2},&\label{pro18k}\\
&\text{The~equality~constraints~in~TABLE~II},&\label{pro18l}
\end{align}\label{pro18}%
\end{subequations}
where $\tilde{\mathcal{U}}=\{$ $C_{k}$, $\bar{C}_{k}$, $\check{C}_{k}$, $\hat{C}_{k}$, $\tilde{C}_{k}$, $\rho$, $\hat{\rho}$, $\tilde{\rho}$, $\bar{\rho}$, $t_{d}$, $\tilde{t}_{d}$, $\hat{t}_{d}$, $\tilde{t}_{u}$, $t_{u}$, $t_{u,k}$, $\tilde{t}_{u,k}$, $t_{d,k}$, $\tilde{t}_{d,k}$, $t_{c,k}$, $\tilde{t}_{c,k}$, $\eta_{k}$, $\tilde{\eta}_{k}$, $\hat{\eta}_{k}$, $\bar{\eta}_{k}$, $\boldsymbol{v}_{\ell}$, $\tilde{\boldsymbol{v}}_{\ell,o_{1}}$, $\boldsymbol{B}_{t,k}$,  $\boldsymbol{B}_{t}$, $\boldsymbol{B}_{r}$, $\boldsymbol{B}_{b}$, $\boldsymbol{B}_{\tilde{t},k}$, $\boldsymbol{B}_{\tilde{t},k^{\prime}}$, $\boldsymbol{F}$, $\tilde{\boldsymbol{F}}\}$.

\section{Proposed PDD-based Algorithm}\label{IV}
Based on the PDD framework, we move the equality constraints
in first six lines of \textbf{TABLE}~\ref{TA2} into the objective function together with
Lagrange multipliers $\lambda_{\hat{C}_{k}}$, $\lambda_{\bar{C}_{k}}$, $\lambda_{\check{C}_{k}}$, $\lambda_{\hat{\rho}}$, $\lambda_{\tilde{\rho}}$, $\lambda_{\bar{\rho}}$, $\lambda_{\check{\rho}}$,  $\lambda_{\tilde{t}_{d}}$, $\lambda_{\hat{t}_{d}}$, $\lambda_{\tilde{t}_{u}}$, $\lambda_{\tilde{t}_{u,k}}$, $\lambda_{\tilde{t}_{d,k}}$, $\lambda_{\tilde{t}_{c,k}}$, $\lambda_{\tilde{\eta}_{k}}$, $\lambda_{\hat{\eta}_{k}}$, $\lambda_{\bar{\eta}_{k}}$,
$\boldsymbol{\lambda}_{\tilde{u}_{k,k^{\prime\prime}}}$, $\boldsymbol{\lambda}_{\boldsymbol{\mu}_{k^{\prime\prime}}}$, $\boldsymbol{\lambda}_{\tilde{\boldsymbol{\mu}}_{k^{\prime\prime}}}$,$\boldsymbol{\lambda}_{\boldsymbol{v}_{b,k}}$,$\boldsymbol{\lambda}_{\tilde{\boldsymbol{v}}_{b,k}}$,$\boldsymbol{\lambda}_{\tilde{\boldsymbol{u}}_{k}}$, $\boldsymbol{\lambda}_{\boldsymbol{w}_{k}}$, $\boldsymbol{\lambda}_{\tilde{\boldsymbol{\omega}}_{k,k^{\prime}}}$, $\boldsymbol{\lambda}_{\boldsymbol{\omega}_{k^{\prime}}}$, $\boldsymbol{\lambda}_{\hat{\boldsymbol{w}}}$, $\boldsymbol{\lambda}_{\tilde{\boldsymbol{w}}}$, $\boldsymbol{\lambda}_{\tilde{\boldsymbol{v}}_{\ell}}$, $\boldsymbol{\lambda}_{\boldsymbol{B}_{\ell}}$ and $\boldsymbol{\lambda}_{\tilde{\boldsymbol{F}}}$ the penalty coefﬁcient $\kappa$. Then the resulting AL problem can be given in (\ref{pro19}).
\begin{figure*}
\begin{subequations}
\begin{align}
\min_{\tilde{\mathcal{U}}}&~t_{u}+t_{d}+\frac{1}{2\kappa}(\|\boldsymbol{F}-\tilde{\boldsymbol{F}}+\kappa\boldsymbol{\lambda}_{\tilde{\boldsymbol{F}}}\|^{2}+|\rho-\hat{\rho}+\kappa\lambda_{\hat{\rho}}|^{2}+|\rho-\tilde{\rho}+\kappa\lambda_{\tilde{\rho}}|^{2}+|\rho-\bar{\rho}+\kappa\lambda_{\bar{\rho}}|^{2}+|\rho-\check{\rho}+\kappa\lambda_{\check{\rho}}|^{2}+|t_{d}-\hat{t}_{d}+\kappa\lambda_{\hat{t}_{d}}|^{2}+\nonumber\\
&|t_{d}-\tilde{t}_{d}+\kappa\lambda_{\tilde{t}_{d}}|^{2}+|t_{u}-\tilde{t}_{u}+\kappa\lambda_{\tilde{t}_{u}}|^{2}+\sum\nolimits_{k=1}^{K}(|C_{k}-\hat{C}_{k}+\kappa\lambda_{C_{k}}|^{2}+|C_{k}-\bar{C}_{k}+\kappa\lambda_{\bar{C}_{k}}|^{2}+|\tilde{C}_{k}-\check{C}_{k}+\kappa\lambda_{\check{C}_{k}}|^{2}+&\nonumber\\
&|t_{u,k}-\tilde{t}_{u,k}+\kappa\lambda_{\tilde{t}_{u,k}}|^{2}+|t_{d,k}-\tilde{t}_{d,k}+\kappa\lambda_{\tilde{t}_{d,k}}|^{2}+|t_{c,k}-\tilde{t}_{c,k}+\kappa\lambda_{\tilde{t}_{c,k}}|^{2}+|\eta_{k}-\tilde{\eta}_{k}+\kappa\lambda_{\tilde{\eta}_{k}}|^{2}+|\hat{\eta}_{k}-\bar{\eta}_{k}+\kappa\lambda_{\bar{\eta}_{k}}|^{2})+&\nonumber\\
&\sum\nolimits_{k^{\prime\prime}=1}^{K}(|\tilde{u}_{k,k^{\prime\prime}}-\tilde{\boldsymbol{u}}_{k}\tilde{\boldsymbol{\mu}}_{k^{\prime\prime}}+\kappa\lambda_{\tilde{u}_{k,k^{\prime\prime}}}|^{2}+\|\boldsymbol{\mu}_{k^{\prime\prime}}-\boldsymbol{B}_{t,k^{\prime\prime}}\tilde{\boldsymbol{w}}_{k^{\prime\prime}}+\kappa\boldsymbol{\lambda}_{\boldsymbol{\mu}_{k^{\prime\prime}}}\|^{2}+\|\tilde{\boldsymbol{\mu}}_{k^{\prime\prime}}-\boldsymbol{B}_{r}^{H}\boldsymbol{\Sigma}_{k}\boldsymbol{\mu}_{k^{\prime\prime}}+\kappa\boldsymbol{\lambda}_{\tilde{\boldsymbol{\mu}}_{k^{\prime\prime}}}\|^{2})
&\nonumber\\
&+\|\boldsymbol{v}_{b,k}-\boldsymbol{B}_{b}\boldsymbol{q}_{k}+\kappa\boldsymbol{\lambda}_{\boldsymbol{v}_{b,k}}\|^{2}+\|\tilde{\boldsymbol{v}}_{b,k}-\boldsymbol{v}_{b,k}^{H}\tilde{\boldsymbol{\Sigma}}\boldsymbol{B}_{t}+\kappa\boldsymbol{\lambda}_{\tilde{\boldsymbol{v}}_{b,k}}\|^{2}+\|\tilde{\boldsymbol{u}}_{k}-\tilde{\boldsymbol{v}}_{b,k}\boldsymbol{F}+\kappa\boldsymbol{\lambda}_{\tilde{\boldsymbol{u}}_{k}}\|^{2}+\|\boldsymbol{w}_{k}-\hat{\boldsymbol{w}}_{k}+\kappa\boldsymbol{\lambda}_{\boldsymbol{w}_{k}}\|^{2}&\nonumber\\
&+\|\tilde{\boldsymbol{w}}_{k}-\bar{\boldsymbol{w}}_{k}+\kappa\boldsymbol{\lambda}_{\tilde{\boldsymbol{w}}_{k}}\|^{2}+\sum\nolimits_{k^{\prime}=1}^{K}(\|\tilde{\boldsymbol{\omega}}_{k,k^{\prime}}-\boldsymbol{B}_{t,k}\tilde{\boldsymbol{\Sigma}}_{k,k^{\prime}}\boldsymbol{\omega}_{k^{\prime}}+\kappa\boldsymbol{\lambda}_{\tilde{\boldsymbol{\omega}}_{k,k^{\prime}}}\|^{2}+\|\boldsymbol{\omega}_{k^{\prime}}-\boldsymbol{B}_{\tilde{t},k}\bar{\boldsymbol{
w}}_{k^{\prime}}+\kappa\boldsymbol{\lambda}_{\boldsymbol{\omega}_{k^{\prime}}}\|^{2})\nonumber\\
&+\sum\nolimits_{o_{1}\neq o_{2}}\|\tilde{\boldsymbol{v}}_{\ell,o_{1}}-(\boldsymbol{v}_{\ell,o_{1}}-\boldsymbol{v}_{\ell,o_{2}})+\kappa\boldsymbol{\lambda}_{\tilde{\boldsymbol{v}}_{\ell,o_{1}}}\|^{2}+\|\boldsymbol{A}(\boldsymbol{v}_{\ell})-\boldsymbol{B}_{\ell}+\kappa\boldsymbol{\lambda}_{\boldsymbol{B}_{\ell}}\|^{2}+\|\boldsymbol{A}(\boldsymbol{v}_{\ell,o})-\boldsymbol{B}_{\ell,o}+\kappa\boldsymbol{\lambda}_{\boldsymbol{B}_{\ell,o}}\|^{2}),&\label{pro19a}\\
\mbox{s.t.}~
&(\ref{pro18b})-(\ref{pro18k}),(\ref{pro21}), (\ref{pro22}),&\label{pro19b}
\end{align}\label{pro19}
\end{subequations} 
\hrulefill
\end{figure*}

\subsection{Proposed SCA-based Algorithm}
We have identified that problem (\ref{pro19}) remains a non-convex problem. Therefore, we focus on the constraints in problem (\ref{pro18}). As mentioned in the references\cite{b28,b29,b30}, we can address these non-convex constraints using the SCA algorithm. Specifically, by employing first-order Taylor series approximation for linearization, we approximate these constraints as convex constraints. Subsequently, we solve the augmented Lagrangian (AL) problem based on the iterative algorithm of the SCA.
The core idea of the SCA algorithm is to convert non-convex constraints into a series of easily solvable convex subproblems through local convexification. This process typically involves performing a first-order Taylor expansion of the non-convex functions, approximating them as linear functions based on the current solution, thus enabling the application of convex optimization techniques.

\begin{figure}
  \centering
  \includegraphics[width=0.45\textwidth, height=0.26\textwidth]{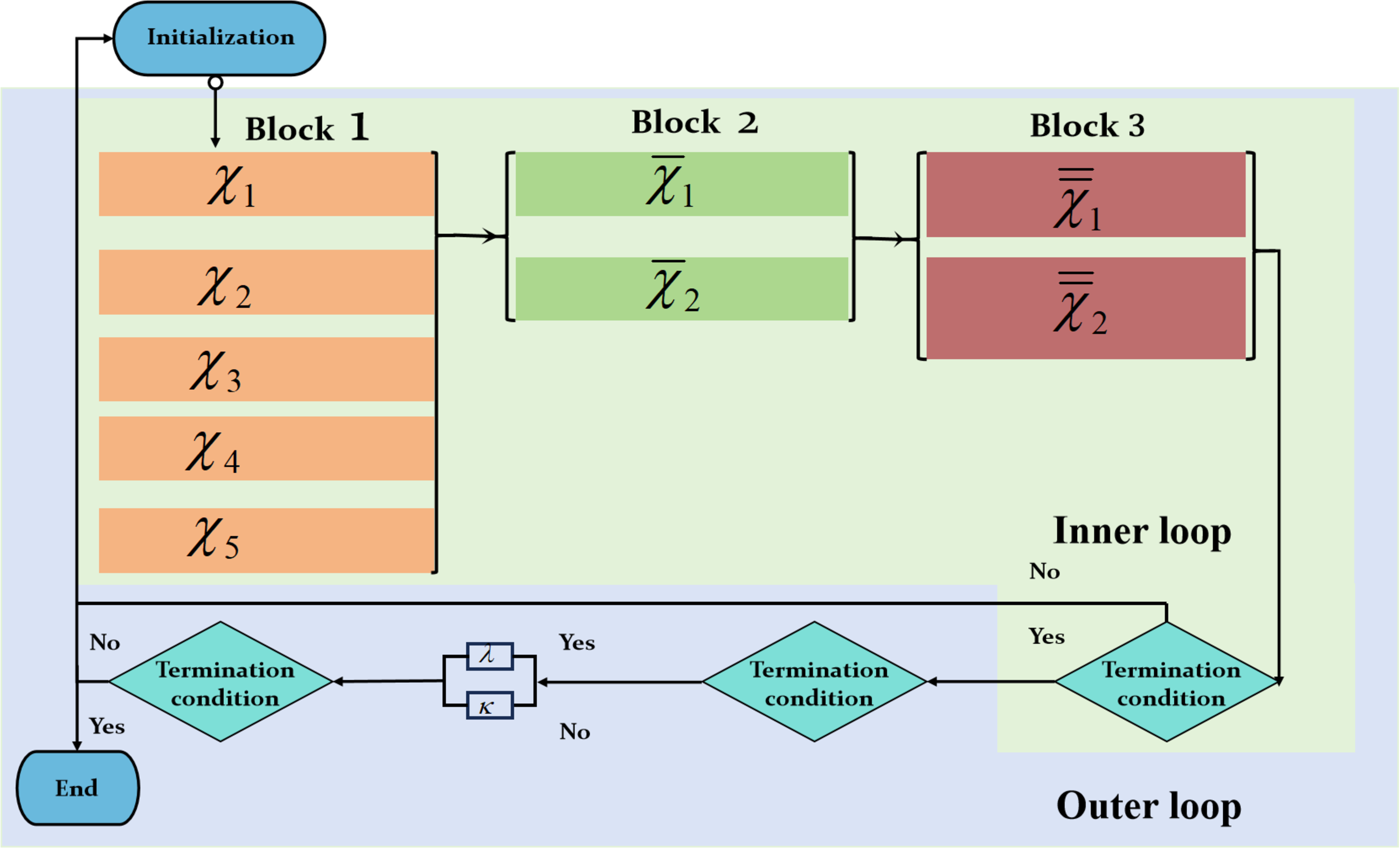}
  \captionsetup{justification=centering}
  \caption{The ﬂow chart of the proposed algorithm.}
\label{FIGURE1}
\end{figure}

First, we use the inequality of arithmetic and geometric means and the Taylor series expansion to deal with constraints  (\ref{pro18b})-(\ref{pro18d}), and they can be rewritten as follows:
\begin{align}
&\bar{\rho}^{2}+L_{a}^{2}/\bar{C}_{k}^{2}\leq 2\tilde{t}_{u,k},
, \hat{\rho}^{2}+L_{a}^{2}/F_{E}^{2}\leq 2\tilde{t}_{d}, \check{\rho}^{2}+L_{a}^{2}/\hat{C}_{k}^{2}\leq 2\tilde{t}_{c,k},\nonumber\\
&(1-\tilde{\rho})^{2}+(\alpha L_{a})^{2}/\check{C}_{k}^{2}\leq 2\tilde{t}_{d,k},\label{pro_20}\\
&C_{k}-\log_{2}(1+\eta_{k}^{(t-1)})+(\eta_{k}-\eta_{k}^{(t-1)})/(2\ln2(1+\eta_{k}^{(t-1)}))\nonumber\\
&\leq 0,~\tilde{C}_{k}-\log_{2}(1+\hat{\eta}_{k}^{(t-1)})+(\hat{\eta}_{k}-\hat{\eta}_{k}^{(t-1)})/(2\ln2(1+\nonumber\\
&\hat{\eta}_{k}^{(t-1)}))\leq 0.\label{pro20}
\end{align}
Then, we use the SCA to deal with the non-convex constraints (\ref{pro18g}) and (\ref{pro18h}). Constraints (\ref{pro18g}) and (\ref{pro18h}) can be rewritten in the following form:
\begin{align}
&\sum_{k^{\prime\prime}\neq k}|\tilde{u}_{k,k^{\prime\prime}}|^{2}+\|\tilde{\boldsymbol{u}}_{k}\|^{2}\sigma^{2}_{r}+\|\boldsymbol{q}_{k}\|^{2}\sigma_{b}^{2}-(\tilde{u}_{k,k}^{(t-1),*}\tilde{u}_{k,k}/\tilde{\eta}_{k}^{(t-1)}+\nonumber\\
&\tilde{u}_{k,k}^{*}\tilde{u}_{k,k}^{(t-1)}/\tilde{\eta}_{k}^{(t-1)}-|\tilde{u}_{k,k}^{(t-1)}|^{2}\tilde{\eta}_{k}/(\tilde{\eta}_{k}^{(t-1)})^{2})\leq 0,\label{pro_21
}\\
&\sum_{k^{\prime}\neq k}\|\tilde{\boldsymbol{\omega}}_{k,k^{\prime}}\|^{2}+\sigma_{b}^{2}-(\tilde{\omega}_{k,k}^{(t-1),*}\tilde{\omega}_{k,k}/\bar{\eta}_{k}^{(t-1)}+\tilde{\omega}_{k,k}^{*}\tilde{\omega}_{k,k}^{(t-1)}/\bar{\eta}_{k}^{(t-1)}\nonumber\\
&-|\tilde{\omega}_{k,k}^{(t-1)}|^{2}\bar{\eta}_{k}/(\bar{\eta}_{k}^{(t-1)})^{2})\leq 0.\label{pro21}
\end{align}
Finally, based on the SCA, we cope with non-convex constraint (\ref{pro21}): 
\begin{align}
&\|\tilde{\boldsymbol{v}}_{\ell,o_{1}}^{(t-1)}\|_{2}-2(\tilde{\boldsymbol{v}}_{\ell,o_{1}}^{(t-1)})^{T}\tilde{\boldsymbol{v}}_{\ell,o_{1}}\geq D,\label{pro22}
\end{align}
where $\tilde{\boldsymbol{v}}_{\ell,o_{2}}^{(t-1)}$ is given as initial MA positions. Let us define the constraint violation $\|\boldsymbol{h}(\mathcal{U}^{(t)})\|_{\infty}$ as: 
\begin{align}
&\|\boldsymbol{h}(\mathcal{U}^{(t)})\|_{\infty}=\max\{|\mathcal{U}_{1}|,\|\boldsymbol{\mathcal{U}}_{1}\|\},~\forall~k,k^{\prime}.\label{pro_18}
\end{align}
Upon executing Algorithm 1 in the inner loop of the proposed PDD-based algorithm, the penalty parameter $\kappa$ is updated using $\kappa^{(t+1)}=c\kappa^{(t)}$, $(0<c< 1)$, depending on the constraint violation condition, and the dual variables are given by:
\begin{align}
&\lambda_{\mathcal{U}_{1}}^{(t+1)}=\lambda_{\mathcal{U}_{1}}^{(t)}+\frac{1}{\kappa^{(t)}}\chi, \text{and}~ \boldsymbol{\lambda}_{\boldsymbol{
\mathcal{U}}_{1}}^{(t+1)}=\boldsymbol{\lambda}_{\boldsymbol{
\mathcal{U}}_{1}}^{(t)}+\frac{1}{\kappa^{(t)}}\boldsymbol{\chi},\label{pro23}
\end{align}
in which $\chi\in\{(\rho-\hat{\rho}),(\rho-\tilde{\rho}),(\rho-\bar{\rho}),(\rho-\check{\rho}),(t_{d}-\hat{t}_{d}),(t_{d}-\tilde{t}_{d}),(t_{u}-\tilde{t}_{u}),(C_{k}-\hat{C}_{k}),(\bar{C}_{k}-\check{C}_{k}), (t_{u,k}-\tilde{t}_{u,k}),(t_{d,k}-\tilde{t}_{d,k}), (t_{c,k}-\tilde{t}_{c,k}),(C_{k}-\tilde{C}_{k}),(\eta_{k}-\tilde{\eta}_{k}),(\hat{\eta}_{k}-\bar{\eta}_{k}),(\tilde{u}_{k,k^{\prime\prime}}-\tilde{\boldsymbol{u}}_{k}\tilde{\boldsymbol{\mu}}_{k^{\prime\prime}})\}$,
$\mathcal{U}_{1}=\{$ $C_{k}$, $\hat{C}_{k}$, $\tilde{C}_{k}$, $\bar{C}_{k}$, $\check{C}_{k}$, $\rho$, $\hat{\rho}$, $\tilde{\rho}$, $\bar{\rho}$, $t_{d}$, $\tilde{t}_{d}$, $\hat{t}_{d}$, $\tilde{t}_{u}$, $t_{u}$, $t_{u,k}$, $\tilde{t}_{u,k}$, $t_{d,k}$, $\tilde{t}_{d,k}$, $t_{c,k}$, $\tilde{t}_{c,k}$, $\eta_{k}$, $\tilde{\eta}_{k}$, $\hat{\eta}_{k}$, $\bar{\eta}_{k}$, $\tilde{u}_{k,k^{\prime\prime}}\}$
and $\boldsymbol{\chi}\in\{(\boldsymbol{F}-\tilde{\boldsymbol{F}})$,$(\boldsymbol{\mu}_{k^{\prime\prime}}-\boldsymbol{B}_{t,k^{\prime\prime}}\tilde{\boldsymbol{w}}_{k^{\prime\prime}})$, $(\tilde{\boldsymbol{\mu}}_{k^{\prime\prime}}-\boldsymbol{B}_{r}^{H}\boldsymbol{\Sigma}_{k}\boldsymbol{\mu}_{k^{\prime\prime}})$,$(\boldsymbol{v}_{b,k}-\boldsymbol{B}_{b}\boldsymbol{q}_{k})$,$(\tilde{\boldsymbol{v}}_{b,k}-\boldsymbol{v}_{b,k}^{H}\tilde{\boldsymbol{\Sigma}}\boldsymbol{B}_{t})$,$(\tilde{\boldsymbol{u}}_{k}-\tilde{\boldsymbol{v}}_{b,k}\boldsymbol{F}),(\boldsymbol{w}_{k}-\hat{\boldsymbol{w}}_{k}), (\tilde{\boldsymbol{w}}_{k}-\bar{\boldsymbol{w}}_{k}), (\tilde{\boldsymbol{\omega}}_{k,k^{\prime}}-\boldsymbol{B}_{t,k}\tilde{\boldsymbol{\Sigma}}_{k,k^{\prime}}\boldsymbol{\omega}_{k^{\prime}})$, $(\boldsymbol{\omega}_{k^{\prime}}-\boldsymbol{B}_{\tilde{t},k}\tilde{\boldsymbol{
w}}_{k^{\prime}})$,$ (\tilde{\boldsymbol{v}}_{\ell,o_{1}}-(\boldsymbol{v}_{\ell,o_{1}}-\boldsymbol{v}_{\ell,o_{2}}))$,$(\boldsymbol{A}(\boldsymbol{v}_{\ell})-\boldsymbol{B}_{\ell})$,$(\boldsymbol{A}(\boldsymbol{v}_{\ell,o})-\boldsymbol{B}_{\ell,o})\}$, $\boldsymbol{\mathcal{U}}_{1}=\{ \boldsymbol{\mu}_{k^{\prime\prime}}$, $\tilde{\boldsymbol{\mu}}_{k^{\prime\prime}}$, $\boldsymbol{v}_{b,k}$, $\tilde{\boldsymbol{v}}_{b,k}$, $\boldsymbol{q}_{k}$, $\tilde{\boldsymbol{u}}_{k}$, $\boldsymbol{w}_{k}$, $\tilde{\boldsymbol{w}}_{k}$, $\hat{\boldsymbol{w}}_{k}$, $\bar{\boldsymbol{w}}_{k}$, $ \tilde{\boldsymbol{v}}_{\ell,o_{1}}$, $\boldsymbol{B}_{\ell,k}$,  $\boldsymbol{B}_{\ell}$, $\boldsymbol{F}$, $\tilde{\boldsymbol{F}}\}$.
$t$ represents the number of outer iterations. The flow of the proposed PDD-based algorithm is outlined in Fig.~\ref{FIGURE1}, based on \cite{b31}. The convergence analysis in \cite{b31} confirms that the PDD-based algorithm for joint hybrid beamforming and resource allocation converges to a set of stationary solutions for problem (\ref{pro15}). The detailed update procedure is next.

\subsection{Variables in \texorpdfstring{$\chi_{1}$}{F},\texorpdfstring{$\chi_{2}$}{F},\texorpdfstring{$\chi_{3}$}{F},\texorpdfstring{$\chi_{4}$}{F},\texorpdfstring{$\chi_{5}$}{F} Update}

\textbf{Optimization~for}~$\chi_{1}=\{\tilde{\boldsymbol{F}},\boldsymbol{w}_{k}\}$\textbf{:}~The subproblem with respect to $\tilde{\boldsymbol{F}}$ is given by:
\begin{subequations}
\begin{align}
\min_{\tilde{\boldsymbol{F}}}&~\|\boldsymbol{F}-\tilde{\boldsymbol{F}}+\kappa\boldsymbol{\lambda}_{\tilde{\boldsymbol{F}}}\|^{2},&\label{pro27a}\\
&\|\tilde{\boldsymbol{F}}\|^{2}\leq P_{r}.&\label{pro27b}
\end{align}\label{pro27}%
\end{subequations}
According to \cite{b27}, problem (\ref{pro27}) can be interpreted as projecting a point onto a spherical surface centered at the origin. Therefore, the solution to problem (\ref{pro27}) is given by:
\begin{align}
\tilde{\boldsymbol{F}}=P_{r}\frac{\boldsymbol{F}+\kappa\boldsymbol{\lambda}_{\tilde{\boldsymbol{F}}}}{\|\boldsymbol{F}+\kappa\boldsymbol{\lambda}_{\tilde{\boldsymbol{F}}}\|+\max(0,P_{r}-\|\boldsymbol{F}+\kappa\boldsymbol{\lambda}_{\tilde{\boldsymbol{F}}}\|)}.\label{pro28}
\end{align}
Similarly, the closed-form solution for the subproblem with regarding to $\boldsymbol{w}_{k}$ and $\tilde{\boldsymbol{w}}_{k}$ can be obtained by using the approach applied in problem (\ref{pro27}).

\textbf{Optimization~for}~$\chi_{2}=\{\boldsymbol{B}_{t,k^{\prime\prime}}$, $\boldsymbol{B}_{t}$, $\boldsymbol{B}_{r}, \boldsymbol{B}_{b}$, $\boldsymbol{B}_{t,k}$,$ \boldsymbol{B}_{\tilde{t},k}$, $\boldsymbol{B}_{\ell}$,$\boldsymbol{B}_{\ell,o}\}$\textbf{:}~
The subproblem with respect to $\boldsymbol{B}_{t,k^{\prime\prime}}$ is given by:
\begin{align}
\min_{|\boldsymbol{B}_{t,k^{\prime\prime}}(l,m_{1})|=1}&\sum\nolimits_{k^{\prime}=1}^{K}\|\boldsymbol{\mu}_{k^{\prime\prime}}-\boldsymbol{B}_{t,k^{\prime\prime}}\tilde{\boldsymbol{w}}_{k^{\prime\prime}}+\kappa\boldsymbol{\lambda}_{\boldsymbol{\mu}_{k^{\prime\prime}}}\|^{2}&\nonumber\\
&+\|\boldsymbol{A}(\boldsymbol{t}_{k^{\prime\prime}})-\boldsymbol{B}_{t,k^{\prime\prime}}+\kappa\boldsymbol{\lambda}_{\boldsymbol{B}_{t,k^{\prime\prime}}}\|^{2}.&\label{pro_23}
\end{align}
According to the Appendix of \cite{b31}, we use one-iteration
 block coordinate descent (BCD) type algorithm, and the key step of updating $\boldsymbol{B}_{t,k}$ in the algorithm is given by: 
 \begin{align}
\boldsymbol{B}_{t,k^{\prime\prime}}(l,m_{1})=b/|b|,\label{pro24}
 \end{align}
where $b=\boldsymbol{\Xi}_{t,k^{\prime}}(l,m_{1})+\tilde{\boldsymbol{\Xi}}_{t,k^{\prime}}(l,m_{1})-\boldsymbol{D}_{t,k^{\prime}}(l,m_{1})$, $\boldsymbol{\Xi}_{t,k^{\prime}}=\tilde{\boldsymbol{w}}_{k^{\prime\prime}}\tilde{\boldsymbol{w}}_{k^{\prime\prime}}^{H}+\boldsymbol{I}$, $\tilde{\boldsymbol{\Xi}}_{t,k^{\prime}}=(\boldsymbol{\mu}_{k^{\prime\prime}}+\kappa\boldsymbol{\lambda}_{\boldsymbol{\mu}_{k^{\prime\prime}}})\tilde{\boldsymbol{w}}_{k^{\prime\prime}}^{H}+\boldsymbol{A}(\boldsymbol{t}_{k^{\prime}})+\kappa\boldsymbol{\lambda}_{\boldsymbol{B}_{t,k^{\prime\prime}}}$, $\boldsymbol{D}_{t,k^{\prime}}=(
\tilde{\boldsymbol{w}}_{k^{\prime\prime}}\tilde{\boldsymbol{w}}_{k^{\prime\prime}}^{H}+\boldsymbol{I})$.
Similarly, we can obtain the solutions of $\boldsymbol{B}_{t}$, $\boldsymbol{B}_{r}, \boldsymbol{B}_{b}$, $\boldsymbol{B}_{t,k}$,$ \boldsymbol{B}_{\tilde{t},k}$, $\boldsymbol{B}_{\ell}$,$\boldsymbol{B}_{\ell,o}$.


\textbf{Optimization~for}~$\chi_{3}=\{\bar{C}_{k}$, $\check{C}_{k}$, $\tilde{t}_{d}$, $\hat{t}_{d}$, $\tilde{t}_{u}$, $\tilde{t}_{u,k}$, $\tilde{t}_{d,k}$, $\tilde{t}_{c,k}$, $\rho\}
$\textbf{:}~
Since this subproblem regarding to $\chi_{3}$ has only convex constraints, it can be solved in closed form using the Lagrange multiplier method. The associated Lagrange function is given in (\ref{pro31}),
\begin{figure*}
\begin{align}
&\mathcal{L}_{1}(\bar{C}_{k})=|C_{k}-\bar{C}_{k}+\kappa\lambda_{\bar{C}_{k}}|^{2}+\kappa_{7,k}(\bar{C}_{k}-\log_{2}(1+\eta_{k}^{(t-1)})+2\ln2(1+\eta_{k}^{(t-1)})/(\eta_{k}-\eta_{k}^{(t-1)})),\label{pro31}
\end{align}
\hrulefill
\end{figure*}
where $\kappa_{1,k}\geq 0$ denotes the Lagrange multiplier  for constraints (\ref{pro_20}) and (\ref{pro20}). By computing the first order optimality condition of $\mathcal{L}_{1}(\chi_{1})$, the optimal value of
 can be derived as (\ref{pro32}) on the next page.
\begin{figure*}
\begin{align}
&\bar{C}_{k}=(C_{k}+\kappa\lambda_{\bar{C}_{k}})-\kappa_{7,k}/2, \kappa_{7,k}^{*}=\left\{0,2((C_{k}+\kappa\lambda_{\bar{C}_{k}})-\log_{2}(1+\eta_{k}^{(t-1)})+2\ln2(1+\eta_{k}^{(t-1)})/(\eta_{k}-\eta_{k}^{(t-1)}))\right\}.\label{pro32}
\end{align}    
\hrulefill
\end{figure*}
Let us denote the optimal $\kappa_{1,k}$ as $\kappa_{1,k}^{*}$, then they are determined
 to fulfill the complementary slackness condition
of (\ref{pro_20}) and (\ref{pro20}), and can be given in (\ref{pro28}) at the top of the next page. The remaining variables of set $\chi_{3}$ can be obtained using the same Lagrange multiplier method.
Employing the same ﬁrst-order optimal method can obtain the optimal solution of the subproblem with respect to $\rho$, which is given in (\ref{pro_33}) at the top of the next page.
\begin{figure*}
\begin{align}
\rho=\left\{\begin{matrix}
(\bar{\rho}+\hat{\rho}+\tilde{\rho}+\check{\rho})-2\kappa(\lambda_{\bar{\rho}}+\lambda_{\hat{\rho}}+\lambda_{\tilde{\rho}}+\lambda_{\check{\rho}})/4,&~\textrm{if}~ 0\leq (\bar{\rho}+\hat{\rho}+\tilde{\rho}+\check{\rho})-2\kappa(\lambda_{\bar{\rho}}+\lambda_{\hat{\rho}}+\lambda_{\tilde{\rho}}+\lambda_{\check{\rho}})/4\leq 1\\
1, &~\textrm{else}~\textrm{if}~\pi(1)<\pi(0)\\
0, &~\textrm{else}\\
\end{matrix}\right.\label{pro_33}
\end{align}
\hrulefill
\end{figure*}

\textbf{Optimization~for}~$\chi_{4}=\{\tilde{\boldsymbol{v}}_{\ell,o_{1}}\}$\textbf{:}~
Similarly, the Lagrange multiplier for the subproblem involving $\tilde{\boldsymbol{v}}_{\ell,o_{1}}$ is given in (\ref{pro29}) at the next page.
\begin{figure*}
\begin{align}
&\mathcal{L}(\tilde{\boldsymbol{v}}_{\ell,o_{1}})=|\tilde{\boldsymbol{v}}_{\ell,o_{1}}-(\boldsymbol{v}_{\ell,o_{1}}-\boldsymbol{v}_{\ell,o_{2}})+\kappa\boldsymbol{\lambda}_{\tilde{\boldsymbol{v}}_{\ell,o_{1}}}|^{2}+\kappa_{2,k}(D-\|\tilde{\boldsymbol{v}}_{\ell,o_{1}}^{(t-1)}\|_{2}+2(\tilde{\boldsymbol{v}}_{\ell,o_{1}}^{(t-1)})^{T}\tilde{\boldsymbol{v}}_{\ell,o_{1}}).\label{pro29}
\end{align}
\hrulefill
\end{figure*}
$\tilde{\boldsymbol{v}}_{\ell,o_{1}}$ is given by:
\begin{align}
&\tilde{\boldsymbol{v}}_{\ell,o_{1}}=\boldsymbol{v}_{\ell,o_{1}}-\boldsymbol{v}_{\ell,o_{1}}-\boldsymbol{\lambda}_{\tilde{\boldsymbol{v}}_{\ell,o_{1}}}+\kappa_{2,k}\tilde{\boldsymbol{v}}_{\ell,o_{1}}^{(t-1)}. \label{pro30}
\end{align}
The optimal Lagrange multiplier $\kappa_{2,k}^{*}$ is in (\ref{pro35}) (next page). 
\begin{figure*}[hbt]
\begin{align}
&\kappa_{5,k}^{*}=\max\left\{0, (\|\tilde{\boldsymbol{v}}_{\ell,o_{1}}^{(t-1)}\|_{2}-D-2(\tilde{\boldsymbol{v}}_{\ell,o_{1}}^{(t-1)})^{T}((\boldsymbol{v}_{\ell,o_{1}}-\boldsymbol{v}_{\ell,o_{2}})-\boldsymbol{\lambda}_{\tilde{\boldsymbol{v}}_{\ell,o_{1}}}))/(2\tilde{\boldsymbol{v}}_{\ell,o_{1}}^{(t-1)})^{T}\tilde{\boldsymbol{v}}_{\ell,o_{1}}^{(t-1)}\right\}. \label{pro35}
\end{align}
\hrulefill
\end{figure*}
\textbf{Optimization~for}~$\chi_{5}=\{\tilde{u}_{k,k^{\prime\prime}},\tilde{\eta}_{k},\tilde{\boldsymbol{u}}_{k}\}$\textbf{:}~
The subproblem with respect to $\chi_{5}$ is given by:
\begin{subequations}
\begin{align}
\min_{\chi_{5}}&~\sum_{k^{\prime}=1}^{K}|\tilde{u}_{k,k^{\prime\prime}}-\tilde{\boldsymbol{u}}_{k}\tilde{\boldsymbol{\mu}}_{k^{\prime\prime}}+\kappa\lambda_{\tilde{u}_{k,k^{\prime\prime}}}|^{2}+|\eta_{k}-\tilde{\eta}_{k}+\kappa\lambda_{\tilde{\eta}_{k}}|^{2}+\nonumber\\
&\|\tilde{\boldsymbol{u}}_{k}-\tilde{\boldsymbol{v}}_{b,k}\boldsymbol{F}+\kappa\boldsymbol{\lambda}_{\tilde{\boldsymbol{u}}_{k}}\|^{2},&\label{pro37a}\\
&(\ref{pro18k}).&\label{pro37b}
\end{align}\label{pro37}%
\end{subequations}
By introducing the Lagrange multipliers $\kappa_{3,k}\geq 0$ and $\kappa_{4,k}\geq 0$ to constraint (\ref{pro37b}), the optimal $\chi_{5}$ is:
\begin{align}
&\tilde{\eta}_{k}(\kappa_{3,k})=(\eta_{k}+\kappa\lambda_{\eta_{k}})-\kappa_{3,k}|\tilde{u}_{k,k}^{(t-1)}|^{2}/(\tilde{\eta}_{k}^{(t-1)})^{2}/2,\nonumber\\
&\tilde{\boldsymbol{u}}_{k}(\kappa_{3,k})=\boldsymbol{\Phi}^{-1}(\kappa_{3,k})\boldsymbol{\phi}(\kappa_{3,k}),\nonumber\\
&\tilde{u}_{k,k^{\prime\prime}}(\kappa_{3,k})=(\tilde{\boldsymbol{u}}_{k}\tilde{\boldsymbol{\mu}}_{k^{\prime\prime}}-\kappa\lambda_{\tilde{u}_{k,k^{\prime\prime}}})/(1+\kappa_{3,k}),~\forall~k\neq k^{\prime},\nonumber\\
&\tilde{u}_{k,k^{\prime\prime}}(\kappa_{3,k})=\tilde{\boldsymbol{u}}_{k}\tilde{\boldsymbol{\mu}}_{k,k}-\kappa\lambda_{\tilde{u}_{k,k}}+\kappa_{3,k}\tilde{u}_{k,k}^{(t-1)}/\tilde{\eta}_{k}^{(t-1)},\label{pro33}
\end{align}
where $\boldsymbol{\Phi}(\kappa_{3,k})=(1+\kappa_{3,k}\sigma_{r}^{2})\boldsymbol{I}+\kappa_{3,k}\tilde{\boldsymbol{\Phi}}_{k}/(1+\kappa_{3,k})$, and $\boldsymbol{\phi}(\kappa_{3,k})=\tilde{\boldsymbol{v}}_{b,k}\boldsymbol{F}-\kappa\lambda_{\tilde{\boldsymbol{u}}_{k}}+\kappa_{3,k}(\sum_{k\neq k^{\prime}}\tilde{\boldsymbol{\mu}}_{k^{\prime\prime}}\kappa\lambda_{\tilde{u}_{k,k^{\prime\prime}}})/(1+\kappa_{3,k})+\kappa_{3,k}\tilde{\boldsymbol{\mu}}_{k,k}\tilde{u}_{k,k}^{(t-1)}/\tilde{\eta}_{k}^{(t-1)}$. We also let $\tilde{\boldsymbol{\Phi}}_{k}=\sum_{k\neq k^{\prime}}\tilde{\boldsymbol{\mu}}_{k^{\prime\prime}}\tilde{\boldsymbol{\mu}}_{k^{\prime\prime}}^{H}$. 
According to\cite{b32}, $\kappa_{3,k}$ must satisfy the complementary slackness condition. Thus:
\begin{align}
&Q_{k}(\chi_{3})=\sum_{k^{\prime}\neq k}|\tilde{u}_{k,k^{\prime\prime}}|^{2}+\|\tilde{\boldsymbol{u}}_{k}\|^{2}\sigma^{2}_{r}+\|\boldsymbol{q}_{k}\|^{2}\sigma_{b}^{2}-(\tilde{u}_{k,k}^{(t-1),*}\nonumber\\
&\tilde{u}_{k,k}/\tilde{\eta}_{k}^{(t-1)}+\tilde{u}_{k,k}^{*}\tilde{u}_{k,k}^{(t-1)}/\tilde{\eta}_{k}^{(t-1)}-|\tilde{u}_{k,k}^{(t-1)}|^{2}\tilde{\eta}_{k}/(\tilde{\eta}_{k}^{(t-1)})^{2})\nonumber\\
&\leq 0.\label{proA1}
\end{align}
When $Q_{k}(\chi_{3}(0))\leq 0$, we have
the optimal $\tilde{u}_{k,k^{\prime\prime}}=\tilde{u}_{k,k^{\prime\prime}}(0)$, $\tilde{\eta}_{k}=\tilde{\eta}_{k}(0)$, and $\tilde{\boldsymbol{u}}_{k}=\tilde{\boldsymbol{u}}_{k}(0)$, otherwise $Q_{k}(\chi_{3})=0$, and it satisfies the following condition:
\begin{align}
&\tilde{\boldsymbol{u}}_{k}^{H}(\kappa_{3,k})\hat{\boldsymbol{\Phi}}(\kappa_{3,k})\tilde{\boldsymbol{u}}_{k}(\kappa_{3,k})-\tilde{\boldsymbol{u}}_{k}^{H}(\kappa_{3,k})\boldsymbol{m}(\kappa_{3,k})-\nonumber\\
&\boldsymbol{m}(\kappa_{3,k})^{H}\tilde{\boldsymbol{u}}_{k}(\kappa_{3,k})+\varpi_{k}(\kappa_{3,k})=0,\label{proA2}
\end{align}
where $\hat{\boldsymbol{\Phi}}(\kappa_{3,k})=\sigma_{r}^{2}\boldsymbol{I}+\tilde{\boldsymbol{\Phi}}_{k}/(1+\kappa_{3,k})^{2}$, $\boldsymbol{m}(\kappa_{3,k})$=$\sum_{k^{\prime}\neq k}\kappa\lambda_{\tilde{u}_{k,k^{\prime\prime}}}^{(t-1)}\tilde{\boldsymbol{\mu}}_{k^{\prime\prime}}/(1+\kappa_{3,k})^{2}$+$\tilde{u}_{k,k}^{(t-1)}/\tilde{\eta}_{k}^{(t-1)}\tilde{\boldsymbol{\mu}}_{k^{\prime\prime}}$, and $\varpi_{k}(\kappa_{3,k})$=$\sum\nolimits_{k^{\prime}\neq k}\kappa^{2}|\lambda_{\tilde{u}_{k^{\prime},k}}|^{2}/(1+\kappa_{3,k})^{2}$+$|\tilde{u}_{k,k}^{(t-1)}|^{2}/(\tilde{\eta}_{k}^{(t-1)})^{2}$+$(\eta_{k}+\kappa\lambda_{\eta_{k}})-\kappa_{3,k}|\tilde{u}_{k,k}^{(t-1)}|^{2}$\\ 
 $/(2\tilde{\eta}_{k}^{(t-1)})^{2}$+$2\mathrm{Re}\{\tilde{u}_{k,k}^{(t-1)}/\tilde{\eta}_{k}^{(t-1)}(\kappa\lambda_{\tilde{u}_{k,k}}+\kappa_{3,k}\tilde{u}_{k,k}^{(t-1)}/\tilde{\eta}_{k}^{(t-1)})$ $\}$.
 For the Hermitian matrix $\tilde{\boldsymbol{\Phi}}_{k}$, use the decomposition:
\begin{align}
\tilde{\boldsymbol{\Phi}}_{k}=\tilde{\boldsymbol{V}}_{k}\boldsymbol{\Lambda}_{k}\tilde{\boldsymbol{V}}_{k}^{H},\label{proA3}
\end{align}
where $\tilde{\boldsymbol{V}}_{k}$ is a unitary matrix which contains the eigenvectors of $\tilde{\boldsymbol{\Phi}}_{k}$, and $\boldsymbol{\Lambda}_{k}$ is a diagonal matrix consisting of the eigenvectors of $\tilde{\boldsymbol{\Phi}}_{k}$. Then $\boldsymbol{\Phi}^{-1}(\kappa_{3,k})$ is denoted as
\begin{align}
&\boldsymbol{\Phi}^{-1}(\kappa_{3,k})=\tilde{\boldsymbol{V}}_{k}((1+\kappa_{3,k}\sigma_{r}^{2})\boldsymbol{I}+\boldsymbol{\Lambda}_{k}\kappa_{3,k}/(1+\kappa_{3,k}))^{-1}\nonumber\\
&\tilde{\boldsymbol{V}}_{k}^{H}.\label{proA4}
\end{align}
Similarly, $\hat{\boldsymbol{\Phi}}(\kappa_{3,k})$ can be rewritten as:
\begin{align}
\hat{\boldsymbol{\Phi}}(\kappa_{3,k})=\tilde{\boldsymbol{V}}_{k}(\sigma_{r}^{2}\boldsymbol{I}+\boldsymbol{\Lambda}_{k}/(1+\kappa_{3,k})^{2})\tilde{\boldsymbol{V}}_{k}^{H}.\label{proA5}
\end{align}
Let $\check{\boldsymbol{\Phi}}(\kappa_{3,k})=(1+\kappa_{3,k}\sigma_{r}^{2})\boldsymbol{I}+\boldsymbol{\Lambda}_{k}\kappa_{3,k}/(1+\kappa_{3,k})$ and $\ddot{\boldsymbol{\Phi}}(\kappa_{3,k})=\sigma_{r}^{2}\boldsymbol{I}+\boldsymbol{\Lambda}_{k}/(1+\kappa_{3,k})^{2}$. Then (\ref{proA2}) is rewritten as:
\begin{align}
&\mathrm{tr}(\check{\boldsymbol{\Phi}}^{-1}\ddot{\boldsymbol{\Phi}}\check{\boldsymbol{\Phi}}^{-1}\tilde{\boldsymbol{V}}_{k}^{H}\boldsymbol{\phi}\boldsymbol{\phi}^{H}\tilde{\boldsymbol{V}}_{k})+\mathrm{tr}(\check{\boldsymbol{\Phi}}^{-1}\tilde{\boldsymbol{V}}_{k}^{H}(\boldsymbol{\phi}\ddot{\boldsymbol{\Phi}}^{H}+\boldsymbol{\phi}^{H}\ddot{\boldsymbol{\Phi}})\tilde{\boldsymbol{V}}_{k})\nonumber\\
&+t_{k}=0.\label{proA6}
\end{align}
Since $\check{\boldsymbol{\Phi}}$ and $\ddot{\boldsymbol{\Phi}}$ are both diagonal matrices, (\ref{proA6}) can be
easily solved using one dimensional search. Finally, by
substituting the optimal $\kappa_{3,k}$, we obtain the solution for
$\chi_{3}$.

\subsection{Variables in \texorpdfstring{$\bar{\chi}_{1}$}{F},\texorpdfstring{$\bar{\chi}_{2}$}{F},\texorpdfstring{$\bar{\chi}_{3}$}{F},\texorpdfstring{$\bar{\chi}_{4}$}{F} update}

\textbf{Optimization~for}~$\bar{\chi}_{1}=\{\boldsymbol{F}$,$\hat{\rho}$,$\tilde{\rho}$,$\bar{\rho}$,$\check{\rho}$,$t_{d}$,$t_{u}$,$t_{u,k}$,$t_{d,k}$,$t_{c,k}$,$C_{k}$,\\
$\tilde{C}_{k}$,$\eta_{k}$,$\hat{\eta}_{k}$,$\boldsymbol{\mu}_{k^{\prime\prime}}$,$\boldsymbol{v}_{b,k}$,$\hat{\boldsymbol{w}}_{k}$,$\bar{\boldsymbol{w}}_{k}\}$\textbf{:}~
The optimal solution of the subproblem involving $\bar{\chi}_{1}$ can be solved by using the Lagrange multipliers method that is adopted in optimization problems for $\chi_{3}$. Specifically, incorporating the objective function and constraints. Once the Lagrangian is established, the first-order optimality conditions iteratively updates $\bar{\chi}_{1}$. Then, the slackness condition is used to derive the updated formula for the Lagrange multipliers, ensuring convergence towards the optimal solution under the given constraints.

\textbf{Optimization~for}~$\bar{\chi}_{2}=\{\boldsymbol{v}_{\ell}\}$\textbf{:}
The subproblem is given by:
\begin{align}
\min_{\boldsymbol{v}_{\ell}}&\sum_{o_{2}}\sum_{\ell}\|\angle\boldsymbol{A}(\boldsymbol{v}_{\ell})-\angle\boldsymbol{B}_{k}+\kappa\boldsymbol{\lambda}_{\boldsymbol{B}_{k}}\|^{2}+\sum_{o_{2}}\|\tilde{\boldsymbol{v}}_{\ell,o_{1}}-(\boldsymbol{v}_{\ell,o_{1}}&\nonumber\\
&-\boldsymbol{v}_{\ell,o_{2}})+\kappa\boldsymbol{\lambda}_{\tilde{\boldsymbol{v}}_{\ell,o_{1}}}\|^{2}&.\label{pro52}
\end{align}
According to (\ref{pro4}) and (\ref{pro5}), problem (\ref{pro52}) can be reformulated as:
\begin{align}
&\mathcal{J}(\boldsymbol{v}_{\ell,o_{1}})=\sum\nolimits_{m_{1}=1}^{N_{u}}(\sum\nolimits_{l=1}^{L}|\frac{2\pi}{\lambda}(\boldsymbol{\pi}_{k}^{l})^{T}\boldsymbol{v}_{\ell,o_{1}}-\angle\boldsymbol{B}_{\ell,o_{1}}|^{2}\nonumber\\
&+\sum\nolimits_{m_{2}=1}^{N_{u}}\|\tilde{\boldsymbol{v}}_{\ell,o_{1}}-(\boldsymbol{v}_{\ell,o_{1}}-\boldsymbol{v}_{\ell,o_{2}})+\kappa\boldsymbol{\lambda}_{\tilde{\boldsymbol{v}}_{\ell,o_{1}}}\|^{2}).\label{pro53}
\end{align}
Based on the first-order optimal condition, $\boldsymbol{v}_{\ell,o_{1}}$ is:
\begin{align}
&(\boldsymbol{I}+\sum_{l=1}^{L}\frac{4\pi^{2}}{\lambda^{2}}\boldsymbol{\pi}_{k}^{l}(\boldsymbol{\pi}_{k}^{l})^{T})\boldsymbol{v}_{\ell,o_{1}}-(\tilde{\boldsymbol{v}}_{\ell,o_{1}}+\sum_{m_{2}=1}^{N_{u}}\boldsymbol{v}_{\ell,o_{2}}+\nonumber\\
&\kappa\boldsymbol{\lambda}_{\tilde{\boldsymbol{v}}_{\ell,o_{1}}})-\sum_{l=1}^{L}\frac{2\pi}{\lambda}\boldsymbol{\pi}_{k}^{l}\angle\boldsymbol{B}_{\ell,o_{1}}=0.\label{pro54}
\end{align}
The solution of (\ref{pro54}) is (\ref{pro56}) (top of this page), where $\mathcal{C}_{t}^{max}$ and $\mathcal{C}_{t}^{min}$ are respectively the maximum and minimum mobile regions of the MA position coordinate $\boldsymbol{v}_{\ell,o_{1}}$. The optimal $\boldsymbol{u}_{r,n_{1}}$,$\boldsymbol{u}_{t,\tilde{n}_{1}}$,$\boldsymbol{r}_{b,j_{1}}$ are obtained using the same approach.
\begin{figure*}
\begin{align}
\boldsymbol{v}_{\ell,o_{1}}=\left\{\begin{matrix}
(\boldsymbol{I}+\sum_{l=1}^{L}\frac{4\pi^{2}}{\lambda^{2}}\boldsymbol{\pi}_{k}^{l}(\boldsymbol{\pi}_{k}^{l})^{T})^{-1}((\tilde{\boldsymbol{v}}_{\ell,o_{1}}+\sum_{m_{2}=1}^{N_{u}}\boldsymbol{v}_{\ell,o_{2}}+\kappa\boldsymbol{\lambda}_{\tilde{\boldsymbol{v}}_{\ell,o_{1}}})+\sum_{l=1}^{L}\frac{2\pi}{\lambda}\boldsymbol{\pi}_{k}^{l}\angle\boldsymbol{B}_{\ell,o_{1}}),&~\textrm{if}~ \boldsymbol{v}_{\ell,o_{1}}\in\mathcal{C}_{t}\\
\mathcal{C}_{k,t}^{max}, &~\textrm{else}~\textrm{if}~J(\mathcal{C}_{k,t}^{max})<J(\mathcal{C}_{k,t}^{min})\\
\mathcal{C}_{k,t}^{min}, &~\textrm{else}\\
\end{matrix}\right.,\label{pro56}
\end{align}
\hrulefill
\end{figure*}
%
%
\subsection{Variables \texorpdfstring{$\bar{\bar{\chi}}_{1}$}{F},\texorpdfstring{$\bar{\bar{\chi}}_{2}$}{F},\texorpdfstring{$\bar{\bar{\chi}}_{3}$}{F},\texorpdfstring{$\bar{\bar{\chi}}_{4}$}{F} update}
%
%
\textbf{Optimization~for}~$\bar{\bar{\chi}}_{1}=\{\tilde{\boldsymbol{w}}_{k},\tilde{\boldsymbol{\mu}}_{k^{\prime\prime}}, \tilde{\boldsymbol{v}}_{b,k},\hat{C}_{k}\}$\textbf{:}
This subproblem is solved by using the Lagrange multipliers method used for optimizing $\chi_{3}$.\\
%
%
\textbf{Optimization~for}~$\bar{\bar{\chi}}_{2}=\{\boldsymbol{q}_{k}\}$\textbf{:}
The subproblem is given by:
\begin{subequations}
\begin{align}
\min_{\boldsymbol{q}_{k}}&~\|\boldsymbol{v}_{b,k}-\boldsymbol{B}_{r}\boldsymbol{q}_{k}+\kappa\boldsymbol{\lambda}_{\boldsymbol{v}_{b,k}}\|^{2}&\\
&(\ref{pro17})&\label{pro_54}
\end{align}
\end{subequations}
It is evident that (\ref{pro_54}) corresponds to projecting a point onto a sphere centered at the origin. In closed-form:
\begin{align}
\boldsymbol{q}_{k}=r\frac{pinv(\boldsymbol{B}_{b})(\kappa\boldsymbol{\lambda}_{\boldsymbol{q}_{k}}-\tilde{\boldsymbol{v}}_{b,k}\boldsymbol{F})}{\|\kappa\boldsymbol{\lambda}_{\boldsymbol{q}_{k}}-\tilde{\boldsymbol{v}}_{b,k}\boldsymbol{F}\|+\max(0,r-\|\kappa\boldsymbol{\lambda}_{\boldsymbol{q}_{k}}-\tilde{\boldsymbol{v}}_{b,k}\boldsymbol{F}\|)},\label{pro55}
\end{align}
where 
$r=\|\boldsymbol{q}_{k}\|^{2}\leq ((\tilde{u}_{k,k}^{(i-1),*}\tilde{u}_{k,k}/\bar{\eta}_{k}^{(i-1)}+\tilde{u}_{k,k}^{*}\tilde{u}_{k,k}^{(i-1)}/\bar{\eta}_{k}^{(i-1)}+|\tilde{u}_{k,k}^{(i-1)}|^{2}\bar{\eta}_{k}/(\bar{\eta}_{k}^{(i-1)})^{2})-\sum_{k^{\prime}\neq k}|\tilde{u}_{k,k^{\prime\prime}}|^{2}-\|\boldsymbol{q}_{k}\|^{2}\sigma_{b}^{2})/\sigma^{2}_{r}$.
\textbf{Optimization~for}~$\bar{\bar{\chi}}_{4}=\{\tilde{\boldsymbol{\omega}}_{k,k^{\prime}},\boldsymbol{\omega}_{k^{\prime}},\bar{\eta}_{k}\}$\textbf{:}
The subproblem with respect to $\bar{\bar{\chi}}_{4}$ is given by:
\begin{subequations}
\begin{align}
\min_{\bar{\bar{\chi}}_{4}}&~\|\tilde{\boldsymbol{\omega}}_{k,k^{\prime}}-\boldsymbol{B}_{t,k}\tilde{\boldsymbol{\Sigma}}_{k,k^{\prime}}\boldsymbol{\omega}_{k^{\prime}}+\kappa\boldsymbol{\lambda}_{\tilde{\boldsymbol{\omega}}_{k,k^{\prime}}}\|^{2}+\|\boldsymbol{\omega}_{k^{\prime}}-\boldsymbol{B}_{\tilde{t},k}\tilde{\boldsymbol{
w}}_{k^{\prime}}\nonumber\\
&+\kappa\boldsymbol{\lambda}_{\boldsymbol{\omega}_{k^{\prime}}}\|^{2}+|\hat{\eta}_{k}-\bar{\eta}_{k}+\kappa\lambda_{\bar{\eta}_{k}}|^{2},&\label{pro60a}\\
&(\ref{pro21}).&\label{pro60b}
\end{align}\label{pro60}%
\end{subequations}
To solve the problem (\ref{pro60}), we can use the same Lagrange multiplier method as applied in problem (\ref{pro37}). 
Finally, the proposed algorithm is summarized in \textbf{Algorithm~\ref{algo1}}
\begin{algorithm}%
\caption{Proposed Algorithm for Problem (\ref{pro14})} \label{algo1}
\hspace*{0.02in}{\bf Initialize:}
$\tilde{\mathcal{U}}^{(0)}$, $\kappa^{(0)}>0$, $\boldsymbol{\lambda}^{(0)}$, $1<c<1$, $t=1$.\\
\hspace*{0.02in}{\bf Repeat:}~$t=t+1$.\\
$\tilde{\mathcal{U}}^{(t)}$ is computed based on problem (\ref{pro15})\\
\hspace*{0.02in}{\bf if:}~$\|\boldsymbol{h}(\tilde{\mathcal{U}}^{t})\|_{\infty}\leq\epsilon_{1}^{(t)}$.\\
Updating $\boldsymbol{\lambda}^{(t+1)}$ based on (\ref{pro20}),
$\kappa^{(t+1)}=\kappa^{(t)}$.\\
\hspace*{0.02in}{\bf else:}~
Updating $\boldsymbol{\lambda}^{(t+1)}=\boldsymbol{\lambda}^{(t)}$,
$\kappa^{(t+1)}=\kappa^{(t)}$.\\
\hspace*{0.02in}{\bf Until:} some termination criterion is met.\\
\end{algorithm}
\subsection{Complexity Analysis}
First focus on the updates of $\chi_{1}$, $\chi_{2}$, $\chi_{3}$, $\chi_{4}$, $\chi_{5}$, $\bar{\chi}_{1}$, $\bar{\chi}_{2}$, $\bar{\bar{\chi}}_{1}$ and $\bar{\bar{\chi}}_{2}$, which are the main determinants of complexity. 
The complexity of updating 
$\chi_{1}$ is $\mathcal{O}(2N_{t}N_{r}+N_{r})$. When updating the $\chi_{2}$, the complexity of one iteration of the proposed BCD-type algorithm is determined by $\mathcal{O}(L^{2}N_{u}^{2}+L^{2}N_{r}^{2}+\tilde{L}^{2}N_{t}^{2}+\tilde{L}^{2}N_{b}^{2})$. The complexity of updating 
$\chi_{3}$ is $\mathcal{O}(5K+3)$. 
The complexity of updating 
\{$\boldsymbol{\mu}_{k}$,$\boldsymbol{v}_{b,k}$\} is $\mathcal{O}(N_{u}^{3}+N_{b}^{3})$. 
The complexity of updating 
$\chi_{4}$ is $\mathcal{O}(2N_{u}+2N_{t}+2N_{r}+2N_{b})$. 
The complexity of updating 
\{$\boldsymbol{\mu}_{k}$,$\boldsymbol{v}_{b,k}$\} is $\mathcal{O}(N_{u}^{3}+N_{b}^{3})$.
The complexity of updating 
$\chi_{5}$ depends on the bisection method used to search for the Lagrange multiplier. The number of iterations required is 
$\mathcal{O}(\log_{2}(\theta_{0,s}/\theta_{s}))$, where 
$\theta_{0,s}$ represents the initial interval size and 
$\theta_{s}$ denotes the tolerance. The complexity of updating 
$\bar{\chi}_{2}$ is $\mathcal{O}(2N_{u}+2N_{t}+2N_{r}+2N_{b})$. The complexity of updating 
$\bar{\bar{\chi}}_{1}$ is $\mathcal{O}(N_{b}^{3}+N_{r}^{3})$. The complexity of updating $\bar{\bar{\chi}}_{2}$ is $\mathcal{O}(N_{u}^{3}+N_{r}^{3}+N_{b}^{3})$. Consequently, the overall complexity of the proposed algorithm can be represented as $\mathcal{O}(T_{1}T_{2}(\log_{2}(\theta_{0,s}/\theta_{s})+2N_{t}N_{r}+5N_{r}+4N_{b}+4N_{u}+4N_{t}+L^{2}N_{u}^{2}+L^{2}N_{r}^{2}+\tilde{L}^{2}N_{t}^{2}+\tilde{L}^{2}N_{b}^{2}+3N_{b}^{3}+2N_{u}^{3}+2N_{r}^{3}+5K+3))$, in which the maximum number of iterations for
the inner and outer loops are denoted by $T_{1}$ and $T_{2}$.

\section{Numerical Results}\label{V}

In the simulation, we assume that the total size of the computational tasks is set to $1\times 10^{7}$ bits for all UEs. The MEC server's computational resources are $1\times 10^{8}$ bits/s, while the local computation capacity for all UEs is $0.4\times 10^{7}$ bits/s. The transmit power $P_{k}$ and the relay transmit power $P_{r}$ are set to $15$~dBm and $30$~dBm, respectively. The noise power levels $\sigma_{u}^{2}=10^{-8}$~W and $\sigma_{b}^{2}=10^{-8}$~W. For the PDD-based algorithm, the tolerance parameter is chosen as  $\epsilon_{1}= 1 \times 10^{-3}$. The initial penalty parameter is set to $\kappa^{0}=2$, with $c= 0.6$. Additionally, we set $\epsilon_{1}= 1\times 10^{-3}$ and $\epsilon_{2}^{m+1}=0.7\epsilon_{2}^{m}$. UEs are uniformly distributed around the BS, with distances randomly distributed within a circular ring of $60$~meters(m) in radius, while interferers are distributed within a circular ring of $30$~m in radius. The BS is located at the center of the ring. We employ the channel model from equations (\ref{pro1}) and (\ref{pro4}), where each UE has the same number of transmission and reception paths, i.e., $L_{k}=2$ and $\tilde{L}=8$. For each UE, 
$\boldsymbol{\Sigma}_{k}=\mathrm\{\sigma_{k,1}, \sigma_{k,2},\ldots,\sigma_{k,L}\}$ and $\tilde{\boldsymbol{\Sigma}}=\mathrm\{\sigma_{1}, \sigma_{2},\ldots,\sigma_{\tilde{L}}\}$ are the diagonal matrix, with each diagonal element following a circularly symmetric complex Gaussian (CSCG) distribution 
$\mathcal{CN}(0, c_{k}^{2}/L)$ and $\mathcal{CN}(0, c^{2}/\tilde{L})$, where $c_{k}^{2}=g_{0}d_{k}^{-\alpha}$ and $c^{2}=g_{0}d_{r}^{-\alpha}$
represent the expected channel power gain for UE $k$ and jammer, $g_{0}$
denotes the expected value of the average channel power gain at a reference distance of $1$~m, and $\alpha$ is the path loss exponent. It should be noted that, for a fair comparison, the total power of the elements in the path-response matrix for UEs with different path numbers is the same, i.e., $\mathbb{E}\{\mathrm{tr}(\boldsymbol{\Sigma}_{k}^{H}\boldsymbol{\Sigma}_{k})\}=c^2_{k}$. To demonstrate the superior performance of the proposed algorithm, this paper compares it with several existing methods. Specifically, methods channel matching (CM)-based, time-division multiple access (TDMA)-based, and local computing algorithms are given in\cite{b33}, while methods alternating position selection (APS), receive MA (RMA), maximum channel
power (MCP) and FPA are given in\cite{b16}.

\begin{figure}[htbp]
\centering
\begin{minipage}[t]{0.48\textwidth}
\centering
\includegraphics[width=0.82\textwidth, height=0.50\textwidth]{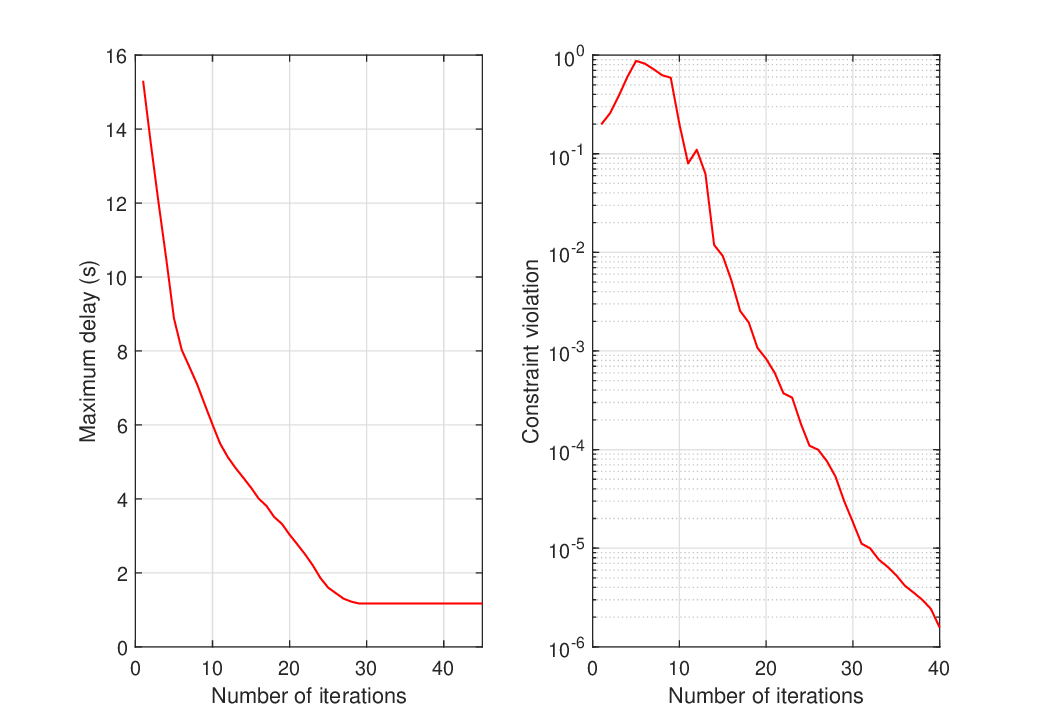}
\put(-150,-5){\small\textbf{(a)}}
\put(-60,-5){\small\textbf{(b)}}
\caption{(a) Maximum latency is plotted against the number of outer-loop iterations. (b) The degree of constraint violation is shown relative to the number of outer-loop iterations.}
\label{FIGURE2}
\end{minipage}
\begin{minipage}[t]{0.48\textwidth}
\centering
\includegraphics[width=0.82\textwidth, height=0.50\textwidth]{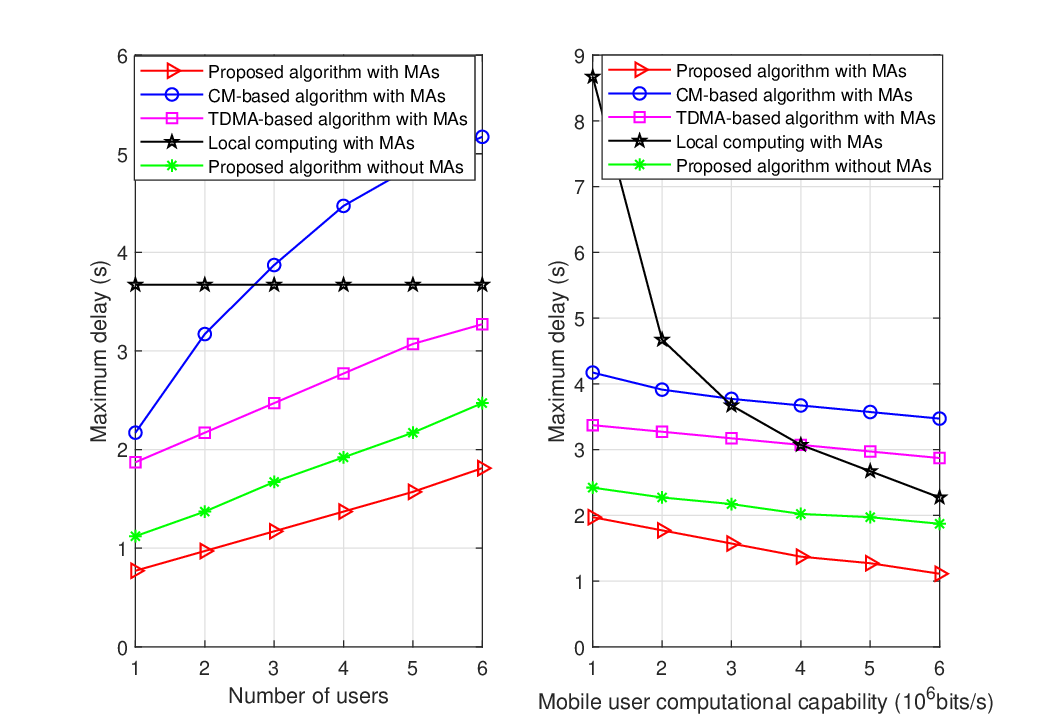}
\put(-150,-5){\small\textbf{(a)}}
\put(-60,-5){\small\textbf{(b)}}
\caption{(a) Maximum latency as a function of the number of UEs. (b) Maximum latency in relation to the UE computational capability.}
\label{FIGURE3}
\end{minipage}
\end{figure}

The relationship between the convergence of the proposed PDD-based joint optimization algorithm and the number of outer-loop iterations is first examined, with simulation parameters set to $N_{t}=2$, $N_{r}=4$, $N_{t}=4$, $N_{b}=8$, $\textrm{SNR}=15$~dB. In Fig.~\ref{FIGURE2}(a), we observe the relationship between the maximum latency obtained by the proposed algorithm and the number of outer-loop iterations. As the number of iterations increases, the system's maximum latency gradually decreases and keeps invariant at around $28$ iterations, indicating the convergence of the proposed algorithm. Fig.~\ref{FIGURE2}(b) illustrates the relationship between the constraint violation metric (i.e., the penalty term (\ref{pro_18})) and the number of iterations. After $30$ iterations, the penalty term decreases to below 10, supporting our conclusion that the proposed algorithm can effectively handle the equality constraints in problem (\ref{pro14}).

In Fig.~\ref{FIGURE3}(a), we analyze the relationship between the number of UEs and the system's maximum latency, comparing our proposed algorithm with other existing algorithms. The simulation parameters are set to 
$N_{t}=2$, $N_{r}=4$, $N_{t}=4$, $N_{b}=8$, $\textrm{SNR}=15$~dB, and all UEs are distributed within a $30$~m ring around the relay station. As shown in Fig.~\ref{FIGURE3}(a), the system's maximum latency increases as the number of UEs rises. This is because, with a limited computational budget, the available resources cannot quickly process each UE's computational tasks as the number of UEs increases, leading to an increase in the system's maximum latency. We then compare the proposed PDD algorithm with existing algorithms based on CM, TDMA, and local computing. The results show that our proposed algorithm exhibits faster performance improvement, demonstrating the advantages of the proposed joint design. Additionally, we observe that the performance of the CM-based algorithm at a distance of $40$~m is even worse than that of the local computing method. This is because the CM algorithm does not fully leverage channel conditions, resource allocation strategies, and MEC server capabilities, whereas the proposed joint design algorithm can provide optimal performance even with a large number of UEs. Fig.~\ref{FIGURE3}(b) illustrates the relationship between the system's maximum latency and the computational capability of mobile UEs. We find that as computational capability increases, the system latency decreases for all four algorithms. Compared to the local computing method, the proposed PDD-based algorithm is less sensitive to UEs' computational capabilities due to the strong computational power of MEC servers. Furthermore, Fig.~\ref{FIGURE3}(a) and Fig.~\ref{FIGURE3}(b) also compare systems without MA, revealing that MA can reduce the system's maximum latency. Therefore, we further analyze the role of MA in reducing the system's maximum latency.

\begin{figure}[htbp]
\centering
\begin{minipage}[t]{0.48\textwidth}
\centering
\includegraphics[width=0.82\textwidth, height=0.50\textwidth]{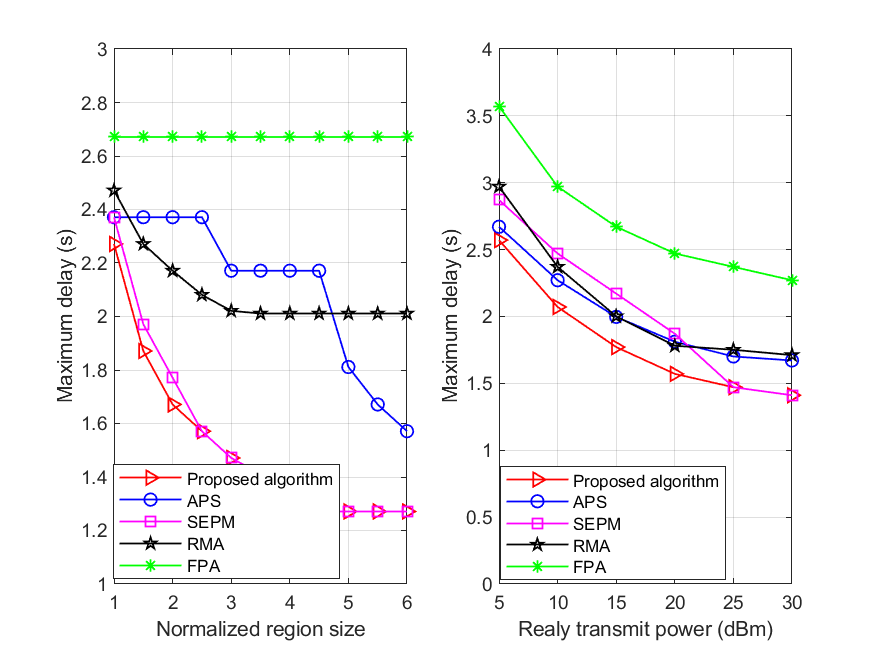}
\put(-150,-5){\small\textbf{(a)}}
\put(-60,-5){\small\textbf{(b)}}
\caption{(a) Maximum latency as a function of the MA mobile region. (b) Maximum latency with respect to the relay transmit power.}
\label{FIGURE4}
\end{minipage}
\begin{minipage}[t]{0.48\textwidth}
\centering
\includegraphics[width=0.82\textwidth, height=0.50\textwidth]{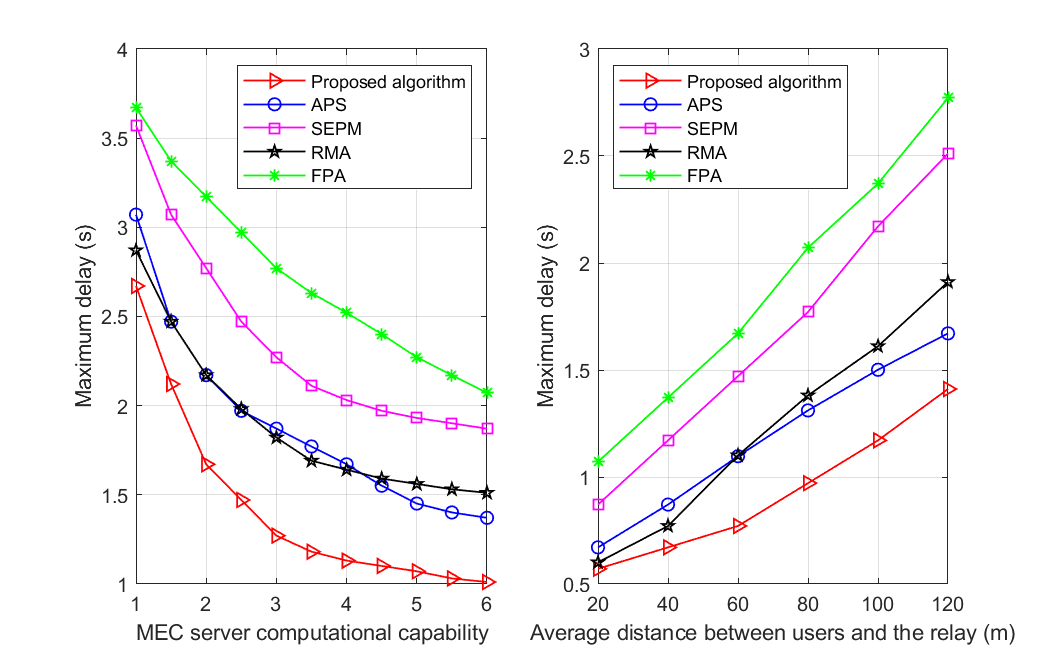}
\put(-150,-5){\small\textbf{(a)}}
\put(-60,-5){\small\textbf{(b)}}
\caption{
(a) Maximum latency in relation to the computational capability of the MEC server. (b) Maximum latency based on the distance between UEs and the BS.}
\label{FIGURE5}
\end{minipage}
\end{figure}

In Fig.~\ref{FIGURE4}(a), the simulation parameters is set as  $N_{t}=2$, $N_{r}=4$, $N_{t}=4$, $N_{b}=8$, $\textrm{SNR}=5$~dB, we compare the relationship between the system's maximum latency and the normalized movable region of the MA for the proposed MA optimization algorithm and the baseline schemes. All schemes outperform the FPA system in reducing system latency and increasing performance gains as the region size expands. Additionally, our proposed algorithm outperforms the RMA scheme, which only considers receiver MA, indicating that the joint optimization of both transmitter and receiver MA provides additional benefits. Compared to the MCP scheme, our proposed method achieves lower latencys even when the MA movable region is small. In Fig.~\ref{FIGURE4}(b), we compare the relationship between the system's maximum latency and the relay transmit power for the proposed MA optimization scheme and the baseline schemes. The results demonstrate that our proposed scheme achieves the best performance. Moreover, an interesting phenomenon is observed: as the relay transmission power gradually increases, the performance of the MCP and the proposed MA schemes converges at a relay power of $25$~dBm. This is because MCP is designed for low SINR conditions and performs poorly in such scenarios. However, as the relay transmit power increases and the system SINR improves, the MCP performance degrades. Considering both Fig.~\ref{FIGURE4}(a) and Fig.~\ref{FIGURE4}(b), we find that our proposed algorithm performs well under both low and high SINR conditions.

Fig.~\ref{FIGURE5}(a) illustrates the relationship between the system's maximum latency and the computational capability of the MEC server, assuming there are four mobile UEs in the system. As the computational capability of the MEC server increases, it can be observed that the system latency for the proposed MA design and other baseline algorithms decreases, due to the availability of more resources on the MEC server. Furthermore, when the MEC's computational capability becomes sufficiently high, the curves for the MA-based and resource allocation algorithms tend to level off, indicating that system latency becomes limited by radio resources and MA DoF. This suggests that once system performance is constrained by radio resources and MA DoF, having redundant MEC computational resources becomes unnecessary.

Fig.~\ref{FIGURE5}(b) shows that as the distance between the UE and the relay increases, the maximum latency for all algorithms increases, with the proposed algorithm demonstrating the best performance. This is because increasing distance deteriorates channel conditions, leading to longer transmission latency for UEs. As the distance between the UE and the relay increases, the performance gap between the MCP-based algorithm and the proposed MA optimization algorithm narrows. This is because, with increased distance, path loss rises and system SINR decreases, under which conditions MCP performs relatively better.

\begin{figure}[htbp]
\centering
\includegraphics[width=0.42\textwidth, height=0.26\textwidth]{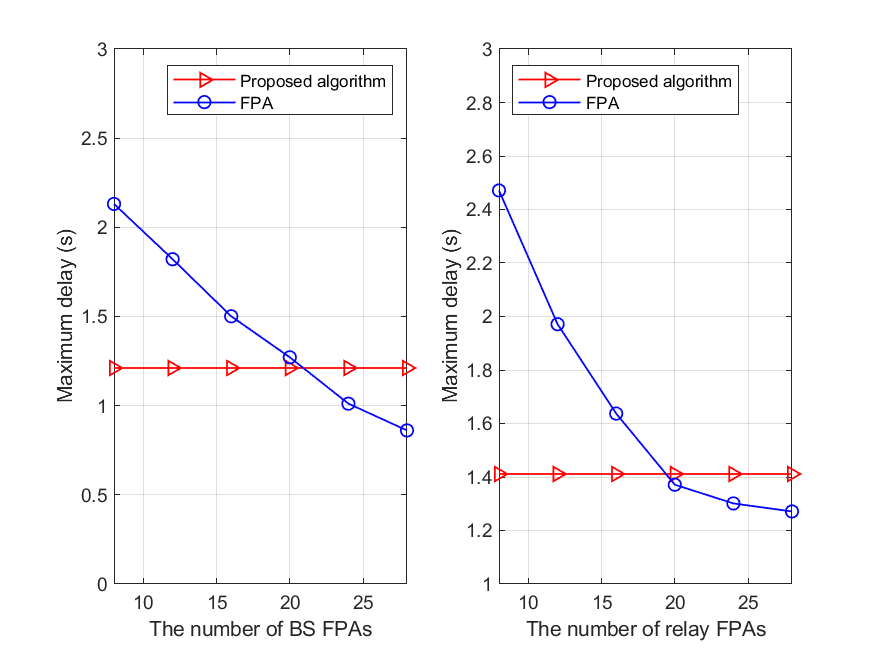}
\put(-150,-5){\small\textbf{(a)}}
\put(-60,-5){\small\textbf{(b)}}
\caption{
(a) Maximum latency about the BS FPA. (b) Maximum latency about the relay FPA.}
\label{FIGURE6}
\end{figure}

From the Fig.~\ref{FIGURE6}(a), it can be observed that as the number of antennas increases, a FPA system requires $N_{b}=22$ antennas to achieve the same latency performance as an MA system with only $N_{b}=8$ antennas. This significant difference underscores the superior efficiency of MAs in terms of latency performance. The advantage of MA over FPA  lies in its ability to dynamically adjust its position and orientation, thereby optimizing spatial diversity and enhancing signal strength in varying conditions. This flexibility allows MA systems to minimize latencys and maintain high performance even in complex communication environments, unlike fixed antennas, which are constrained by their static positioning and limited adaptability. Moreover, the inherent scalability and efficiency of MA systems make them highly suitable for real-world deployments. MA systems reduce hardware and spatial resource demands by requiring fewer antennas to achieve comparable or superior performance, enabling more cost-effective and practical deployment in 6G communication networks. 

Similarly, in the Fig.~\ref{FIGURE6}(b), it can be observed that as the number of relay antennas increases, a relay system equipped with fixed antennas requires $N_{t}=N_{r}=20$ antennas to achieve the same latency performance as a relay system with only $N_{t}=N_{r}=4$ MAs. This highlights the significant efficiency advantage of MAs over fixed antennas in terms of latency performance. The superiority of MA lies in its ability to dynamically adjust its position and orientation to optimize the communication link, effectively leveraging spatial diversity and adapting to varying network and environmental conditions. The results also show the use of MAs can significantly reduce the total number of antennas needed, which directly translates into a reduction in hardware costs for relay-aided systems.

\section{Conclusion}\label{VI}
This paper proposes an efficient joint optimization algorithm for MA positioning and resource allocation for a novel MA-enabled relay-assisted D2D MEC system. The system latency minimization problem is reformulated through a series of appropriate transformations. Subsequently, based on the SCA and PDD methods, we develop an innovative dual-loop distributed joint optimization algorithm for the MA positioning and beamforming at the UE, relay, and BS, as well as task offloading rates and resource allocation at the MEC server. Our simulation results demonstrate the superiority of the joint design and the benefits of integrating MA with relay-assisted MEC systems. The simulation results also show that MA technology in relay-assisted MEC D2D systems can significantly reduce the required number of antennas while ensuring system latency performance. This advantage not only optimizes the system architecture but also effectively lowers costs, thereby enhancing the scalability of relay-assisted MEC D2D systems. This is of great significance for the massive deployment of future 6G networks, as it not only improves network efficiency but also better accommodates the high density, meeting the growing demand for bandwidth and low-latency requirements.



\end{document}